\documentclass[11pt,a4paper,useAMS,usenatbib]{emulateapj}
\usepackage{epsfig}
\usepackage{amsmath}
\usepackage{natbib}

\begin{document}

\title{Hot gaseous coronae around spiral galaxies: \\ probing the Illustris Simulation}

\author{\'Akos Bogd\'an\altaffilmark{1}, Mark Vogelsberger\altaffilmark{2}, Ralph P. Kraft\altaffilmark{1}, \\ Lars Hernquist\altaffilmark{1}, Marat Gilfanov\altaffilmark{3,4}, Paul Torrey\altaffilmark{2,5}, Eugene Churazov\altaffilmark{3,4}, Shy Genel\altaffilmark{1}, \\ William R. Forman\altaffilmark{1},  Stephen S. Murray\altaffilmark{1,6}, Alexey Vikhlinin\altaffilmark{1}, Christine Jones\altaffilmark{1},  and Hans B\"ohringer\altaffilmark{7}} 

\affil{\altaffilmark{1} Harvard-Smithsonian Center for Astrophysics, 60 Garden Street, Cambridge, MA 02138, USA}
\affil{\altaffilmark{2} Department of Physics, Kavli Institute for Astrophysics and Space Research, Massachusetts Institute of Technology, Cambridge, MA 02139, USA}
\affil{\altaffilmark{3} Max-Planck-Institut f\"ur Astrophysik, Karl-Schwarzschild-str. 1, 85748 Garching, Germany}
\affil{\altaffilmark{4} Space Research Institute, Russian Academy of Sciences, Profsoyuznaya 84/32, 117997 Moscow, Russia}
\affil{\altaffilmark{5} TAPIR, Mailcode 350-17, California Institute of Technology, Pasadena, CA 91125, USA}
\affil{\altaffilmark{6} Department of Physics and Astronomy, Johns Hopkins University, 3400 North Charles Street, Baltimore, MD 21218, USA}
\affil{\altaffilmark{7} Max Planck Institute for extraterrestrial Physics, Giessenbachstra§e 1, 85748 Garching, Germany}

\email{E-mail: abogdan@cfa.harvard.edu}

\shorttitle{HOT X-RAY CORONAE AROUND MASSIVE SPIRALS}
\shortauthors{BOGD\'AN ET AL.}

\begin{abstract}
The presence of hot gaseous coronae around present-day massive spiral galaxies is a fundamental prediction of galaxy formation models. However, our observational knowledge remains  scarce, since to date only four gaseous coronae were detected around spirals with massive stellar bodies ($\gtrsim2\times10^{11} \ \rm{M_{\odot}}$). To explore the hot coronae around lower mass spiral galaxies, we utilized \textit{Chandra} X-ray observations of a sample of eight normal spiral galaxies with stellar masses of $(0.7-2.0)\times10^{11} \ \rm{M_{\odot}}$. Although statistically significant diffuse X-ray emission is not detected beyond the optical radii ($\sim$$20$ kpc) of the galaxies, we derive $3\sigma$ limits on the characteristics of the coronae. These limits, complemented with previous detections of NGC~1961 and NGC~6753, are used to probe the Illustris Simulation. The observed $3\sigma$ upper limits on the X-ray luminosities and gas masses exceed or are at the upper end of the model predictions. For NGC~1961 and NGC~6753 the observed gas temperatures, metal abundances, and electron density profiles broadly agree with those predicted by Illustris. These results hint that the physics modules of Illustris are broadly consistent with the observed properties of hot coronae around spiral galaxies. However, a shortcoming of Illustris is that massive black holes, mostly residing in giant ellipticals, give rise to powerful radio-mode AGN feedback, which results in under luminous coronae for ellipticals. 
\end{abstract}


\section{Introduction}
\label{sec:introduction}
The existence of gaseous X-ray coronae in the dark matter halos of massive galaxies is predicted by galaxy formation models \citep[e.g.][]{white78,white91,toft02,rasmussen09,crain10,vogelsberger13,marinacci14}.  Although simulations agree that these hot coronae should be ubiquitous, various models made very different predictions.  The disagreement among models originates from two sources. First, the incomplete realization of the physical processes may result in different galaxy populations. Specifically, many earlier models lacked efficient supernova and/or AGN feedback, which often yielded overly massive stellar components relative to their dark matter halos due to the strong gas cooling. Second, using different hydrodynamical schemes resulted in large systematic differences in the global state of baryons \citep{vogelsberger12}. Thus, the predicted properties of hot coronae also sensitively depend on the applied hydrodynamical scheme \citep{bogdan13a}. The ever increasing computing power plays a major role to overcome the issues related to incomplete modeling of the physical processes. Indeed,  state-of-the-art simulations  are now capable of studying large cosmological volumes combined with high numerical resolution, which simultaneously allows the study of large galaxy samples and robust modeling of complex physical processes \citep[e.g.][]{vogelsberger14a}. Additionally, detailed tests helped to understand the differences between the recently developed moving mesh and the traditional smoothed particle hydrodynamics codes.

\begin{table*}
\caption{The properties of the sample galaxies.}
\begin{minipage}{18.5cm}
\renewcommand{\arraystretch}{1.3}
\centering
\begin{tabular}{c c c c c c c c c c c}
\hline 
Name &  Distance & $1\arcmin$ scale  &  $N_{\rm H}$ & $L_{K} $    &  $M_{\rm \star}/L_{\rm K} $            &  $M_{\rm \star}$ & SFR & Morph.    & $r_{\rm 200}$     & $D_{\rm 25}$        \\ 
   &   (Mpc)    & (kpc) & (cm$^{-2}$) &   ($\rm{L_{K,\odot}}$) & ($\rm{M_{\odot}/L_{K,\odot}}$) & ($\rm{M_{\odot}}$) &($\rm{M_{\odot} \ yr^{-1}} $) &  type       & (kpc)   & (')            \\ 
     &   (1)    &            (2)           &   (3)      &     (4)     &      (5)                 &   (6) &  (7)                  &   (8)     & (9)      & (10)       \\
\hline 
NGC~266  & $ 60.3  $ & 17.55 & $  5.7 \times 10^{20}$ & $ 2.5 \times 10^{11}$ & 0.80 & $ 2.0 \times 10^{11}$  &  2.4 & Sab & 410 & 2.95 \\
NGC~1097  & $ 14.2  $ & 4.12 & $  2.0 \times 10^{20}$ & $ 1.3 \times 10^{11}$ & 0.77 & $ 1.0 \times 10^{11}$  &  5.8 & SBb & 238  & 10.47 \\
NGC~2841  &  $ 14.1 $ &  4.09 & $  1.3 \times 10^{20}$ & $  1.5 \times 10^{11}$& 0.78 &  $ 1.2 \times 10^{11}$ &  0.8 & Sb & 363 &6.92  \\
NGC~5005  & $ 18.5  $ & 5.38 & $  1.1 \times 10^{20}$ & $ 1.9 \times 10^{11}$ & 0.77 & $ 1.4 \times 10^{11}$  &  4.9 & SABb & 273  & 4.79 \\
NGC~5170  & $ 21.6  $ & 6.29 & $  6.7 \times 10^{20}$ & $ 8.5 \times 10^{10}$ & 0.83 & $ 7.1 \times 10^{10}$  &  0.4 & Sc & 269 & 7.94 \\
NGC~5529  & $ 36.0  $ & 10.45 & $  1.0 \times 10^{20}$ & $ 1.0 \times 10^{11}$ & 0.75 & $ 7.8 \times 10^{10}$  &  2.0 & Sc & 314 & 5.75 \\
NGC~5746  & $ 19.8  $ & 5.79 & $  3.4 \times 10^{20}$ & $ 1.4 \times 10^{11}$ & 0.77 & $ 1.1 \times 10^{11}$  &  0.6 & SABb & 353 & 7.24 \\
ESO~445-081&$ 69.3$ & 20.16 & $  5.0 \times 10^{20}$ & $ 1.2 \times 10^{11}$ & 0.78 & $ 9.4 \times 10^{10}$  &  5.6 & SBbc & 590  & 1.70  \\
\hline \\
\end{tabular} 
\end{minipage}
\textit{Note.} Columns are as follows. (1) Distance taken from \citet{tully09} (2) $1\arcmin$ scale at the adopted distance. (3)  Line-of-sight column density provided by the LAB survey \citep{kalberla05}. (4) Total K-band luminosity from 2MASS. (5) K-band mass-to-light ratios computed from \citet{bell03} using the $B-V$ color indices of galaxies. (6) Total stellar mass based on the K-band luminosity and the K-band mass-to-light ratios. (7) Star formation rate computed from the \textit{IRAS} $60 \ \rm{\mu m}$ and $100 \ \rm{\mu m}$ flux -- for details see Section \ref{sec:fir}. (8) Morphological type, taken from HyperLeda. (9) Virial radius of the galaxies, estimated from the maximum rotation velocity following \citet{bogdan13a}. (10) Major axis diameter of the $D_{25}$  ellipse.  \\

\label{tab:list1}
\end{table*}  

Although hot X-ray coronae are of fundamental importance, their detection and characterization poses major observational challenges around spiral galaxies, which system offer a clean test of galaxy formation models \citep{bogdan13a}. The main difficulty is due to the faint but extended nature of the coronae around spirals, which in turn results in low signal-to-noise ratios. Additionally, the brightest X-ray coronae are associated with the most massive galaxies, which systems are rare in the local Universe ($D\lesssim100$ Mpc). Despite these challenges, numerous observational studies aimed to explore hot gaseous coronae around nearby spiral galaxies, which systems offer a clean test of galaxy formation models \citep[e.g.][]{benson00,rasmussen09}. However, the first X-ray coronae around spirals have only recently been detected using \textit{Chandra}, \textit{XMM-Newton}, and \textit{ROSAT} data, which revealed gaseous coronae around NGC~1961, UGC~12591, NGC~6753, and NGC~266 \citep{anderson11,dai12,bogdan13a,bogdan13b}. Due to the large distances of these galaxies and/or the relatively shallow X-ray observations, the detailed characterization of the coronae was only possible  for NGC~1961 and NGC~6753 \citep{anderson11,bogdan13a}.

Despite these advances, the presently available sample is still scarce and only includes massive galaxies ($M_{\rm \star}\gtrsim2\times10^{11} \ \rm{M_{\odot}}$), which presents limitations in using the hot coronae as probes of galaxy formation models. While extra-planar diffuse emission has been detected around several lower mass spiral galaxies \citep{li06,li11,bogdan08}, this gaseous emission cannot be attributed to the hot corona of infalling gas. Indeed, the diffuse gas in these galaxies do not extend to large radii, and exhibits a bipolar morphology. The morphology combined with the mass and energy budget of the galaxies hint that the gas is outflowing and the outflows are sustained by the energy input from type Ia supernovae. Additionally, several studies have explored the presence of diffuse emission within the central regions of spiral galaxies \citep{li13,li14}. Although a certain fraction of this emission may originate from the infalling hot gas, it is nearly impossible to disentangle this emission from other diffuse X-ray emitting components, in particular from the hot gas associated with star-formation. Therefore, it is desirable to probe hot coronae \textit{beyond} the optical radii of the galaxies. Exploring extended X-ray coronae around lower mass spirals will allow us to place stronger constraints on galaxy formation simulations and to probe a broader galaxy population. However, due to their faint nature, the X-ray coronae may remain undetected around  galaxies with $M_{\rm \star}<2\times10^{11} \ \rm{M_{\odot}}$. While non-detections only allow limits on various properties of the coronae, these limits provide invaluable input to probe galaxy formation models. Thus, the goal of this work is to extend the presently available sample with a focus on lower mass spiral galaxies. 

\begin{figure*}
  \begin{center}
    \leavevmode
      \epsfxsize=18cm\epsfbox{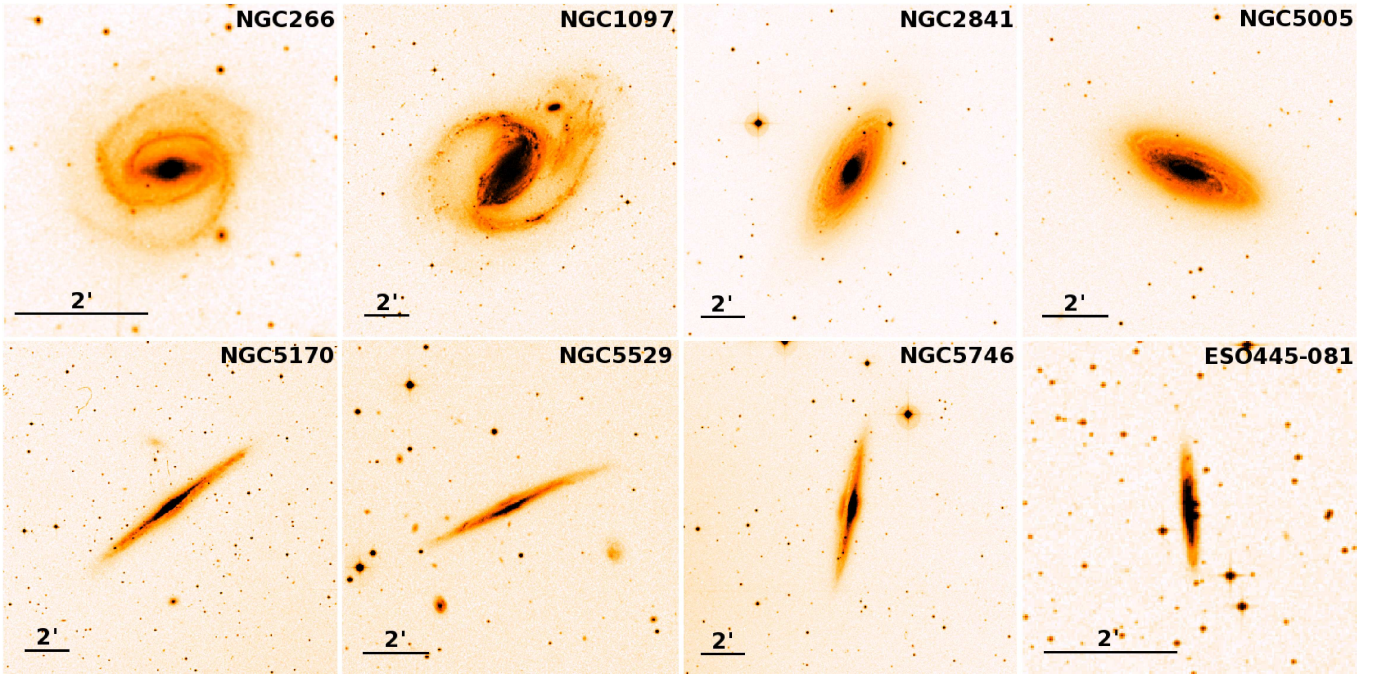}
      \caption{DSS B-band images of the spiral galaxies in our sample. All galaxies reveal late-type morphologies and have a prominent disk component. The images show that the selected galaxies are relatively undisturbed, only NGC~1097 may be experiencing a minor merger.  The sample galaxies exhibit a wide range of inclinations, including face-on and edge-on systems as well.}
\vspace{0.5cm}
     \label{fig:dss}
  \end{center}
\end{figure*}

The Illustris Simulation, which is based on the moving-mesh code \textsc{arepo} \citep{springel10}, offers the ideal framework to compare the observational results with  those predicted by a modern hydrodynamical cosmological simulation. Illustris includes the main physical processes that influence galaxy evolution, such as primordial and metal line cooling, star-formation, chemical enrichment, energetic feedback from supernovae and AGN \citep{vogelsberger14b}. Therefore, Illustris results in a realistic galaxy population of elliptical and spiral galaxies \citep{vogelsberger13}. Additionally, Illustris was performed in a large cosmological box, which consists of a broad population of spiral galaxies with $M_{\rm \star}\gtrsim10^{11} \ \rm{M_{\odot}}$, allowing a comparison with observational results. Thus, the goal of our study is to probe hot X-ray coronae around spiral galaxies and compare the observed properties with those predicted by Illustris. 

The paper is structured as follows. In Section 2 we introduce the observed and simulated galaxy sample. In Section 3 the data reduction is described. We present the results in Section 4, where we discuss the constraints on the characteristics of the hot X-ray coronae. In Section 5 we compare the observed and predicted properties of the coronae. We discuss our results in Section 6 and summarize in Section 7.

\section{Galaxy samples}
\subsection{The observed galaxy sample}
\label{sec:galaxies}
To probe the hot X-ray coronae around spiral galaxies, we utilize \textit{Chandra} X-ray observations. While \textit{XMM-Newton} offers larger effective area, a particular advantage of \textit{Chandra} is its lower and more stable instrumental background level, which makes it ideal to study the faint X-ray emission  expected from hot coronae around spiral galaxies with $M_{\rm \star}<2\times 10^{11} \ \rm{M_{\odot}}$. 

To this end, we searched for spiral galaxies with publicly available \textit{Chandra} observations. To ascertain the morphological type of the galaxies, we relied on the HyperLeda data base and included Sa-Sc systems. We further filtered the sample by selecting galaxies in the distance range of $14<D<40$ Mpc. The lower distance boundary assures that at least the inner regions of the coronae can fit in the \textit{Chandra} field-of-view, while the higher boundary was introduced to maximize the signal-to-noise ratio. We also required that the sample galaxies be sufficiently massive ($M_{\rm \star}>7\times10^{10} \ \rm{M_{\odot}}$), which ensures that they reside in a massive dark matter halo, and host a more luminous and hotter X-ray corona. Additionally, we excluded those galaxies, which are located in rich galaxy groups/clusters, and only retained field galaxies or systems residing in poor groups with $<20$ members. This assures that we can probe the hot corona associated with the galaxy and not a larger-scale group/cluster atmosphere. Moreover, we excluded galaxies undergoing a significant merger event, since the dark matter halos, and hence the hot gaseous coronae, of these galaxies may be disturbed. Finally, we removed starburst galaxies (for details see Section \ref{sec:disc}) and systems with high star-formation rates ($>10 \ \rm{M_{\odot} \ yr^{-1}}$), since these galaxies could be prone to starburst driven winds. Based on these criteria we identified six nearby spiral galaxies. 

This sample was further expanded by two additional spiral galaxies, NGC~266 and ESO~445-081. While these galaxies are  more distant, their  \textit{ROSAT} observations indicate the presence of luminous X-ray coronae \citep{bogdan13b}. Thus, we utilized follow-up \textit{Chandra} observations to further study these galaxies. 

Figure \ref{fig:dss} shows the Digitized Sky Survey (DSS) B-band images of the 8 spiral galaxies in our sample. While all galaxies reveal late-type morphologies, they exhibit a wide range of inclinations from face-on to edge-on systems. The physical properties of these spirals are listed in Table \ref{tab:list1}.

\begin{table}
\caption{The list of analyzed \textit{Chandra} observations.}
\begin{minipage}{8.75cm}
\renewcommand{\arraystretch}{1.3}
\centering
\begin{tabular}{c c c c c }
\hline 
Galaxy & Obs ID &  $T_{\rm{obs}}$ (ks) & $T_{\rm{filt}}$ (ks) & Instrument \\
\hline
NGC~266 & 16013$^{\dagger}$ & 85.0 &  85.0 & ACIS-I \\
NGC~266 & 16301   & 41.5 &  41.5 & ACIS-I \\

NGC~1097 & 1611$^{\dagger}$ & 5.4 & 5.0 & ACIS-S \\
NGC~1097 & 2339 & 5.7 & 5.0 & ACIS-S \\

NGC~2841 & 389 & 1.8 & 1.8 & ACIS-S \\
NGC~2841 & 6096$^{\dagger}$ & 28.2 & 21.3 & ACIS-S \\

NGC~5005 & 4021 & 4.9 & 4.7 & ACIS-S \\

NGC~5170 & 3928 & 33.0 & 33.0 & ACIS-I \\

NGC~5529 & 4163   & 89.2  & 71.0 & ACIS-I \\
NGC~5529 & 12255 & 60.4  & 50.8 & ACIS-S \\
NGC~5529 & 12256$^{\dagger}$ & 118.5& 96.0 & ACIS-S \\
NGC~5529 & 13118 & 44.6  & 38.3 & ACIS-S \\
NGC~5529 & 13119 & 54.3  & 48.0 & ACIS-S \\

NGC~5746 & 4021 & 36.8 & 33.6 & ACIS-I \\

ESO~445-081 & 16302 & 39.2 & 39.2 & ACIS-I \\
\hline \\
\end{tabular} 
\end{minipage}
$^{\dagger}$ The coordinate system of these observations  was used as reference when merging observations.   \\
\label{tab:list2}
\end{table}

\subsection{The simulated galaxy sample}
To compare the observed characteristics of the coronae with those predicted by Illustris, we used two approaches. First, we selected all galaxies independently of their morphological types, whose stellar mass is in the range of  $(0.2-20)\times 10^{11}  \ \rm{M_{\odot}}$. Due to their shallow potential well, the dominant fraction of galaxies with $M_{\rm \star}<2\times 10^{10}  \ \rm{M_{\odot}}$ cannot confine a significant amount of hot X-ray gas. This approach offers a statistically significant galaxy sample, allowing us to characterize the gaseous coronae from low stellar mass systems to the most massive galaxies. Additionally, this large galaxy sample removes the distorting effects of outliers, which ensures that we can explore the general trends in the simulated properties of the hot coronae. 

Second, we analyzed a sub-sample of spiral galaxies that were selected by \citet{vogelsberger14a}. These 42 ``textbook'' spirals are sufficiently massive and are disk dominated systems. Therefore, the ``textbook'' spirals are the analogs of the observed galaxy sample, and hence offer an ideal 
 basis to confront the \textit{Chandra} observations with the Illustris Simulation.

\begin{table*}
\caption{Constraints on the parameters of the diffuse hot gaseous coronae.}
\begin{minipage}{18cm}
\renewcommand{\arraystretch}{1.3}
\centering
\begin{tabular}{c c c c c c c c}
\hline 
Galaxy& Radial range & $L_{\rm{0.5-2keV,abs}}$ & $L_{\rm{bol}}$ & $M_{\rm{gas}}$ & $n_{\rm{e}}$  & $t_{\rm{cool}}$ & $\dot M$\\ 
 & $r_{200}$ & $\rm{erg \ s^{-1}}$ & $\rm{erg \ s^{-1}}$ & $\rm{M_{\odot}}$ & $\rm{cm^{-3}}$ & Gyr & $\rm{M_{\odot} \ yr^{-1}}$ \\ 
\hline
NGC~266  & 0.05-0.15 & $<1.9 \times 10^{40}$ &  $<8.9 \times 10^{40}$  & $<1.9\times10^{10}$& $<8.0\times10^{-4}$ &$>2.9$ & $<6.5$ \\
NGC~1097& 0.05-0.15 & $<1.5 \times 10^{39}$ &  $<6.1 \times 10^{39}$  & $<2.2\times10^{9}$ &$<5.0\times10^{-4}$ & $>4.6$ & $<0.5$ \\
NGC~2841& 0.05-0.15 & $<2.9 \times 10^{39}$ &  $<1.1 \times 10^{40}$  & $<5.8\times10^{9}$ &$<3.6\times10^{-4}$ & $>6.4$ & $<0.9$\\
NGC~5005& 0.05-0.15 & $<2.3 \times 10^{39}$ &  $<8.7 \times 10^{39}$  & $<3.3\times10^{9}$ & $<4.9\times10^{-4}$ & $>4.7$ & $<0.7$\\
NGC~5170&0.05-0.15  & $<3.0 \times 10^{39}$ &  $<1.6 \times 10^{40}$  & $<4.3\times10^{9}$ & $<6.6\times10^{-4}$ & $>3.5$ & $<1.2$\\
NGC~5529& 0.05-0.15 & $<4.7 \times 10^{39}$ &  $<1.8 \times 10^{40}$  & $<5.8\times10^{9}$ &$<5.6\times10^{-4}$ & $>4.1$ & $<1.4$\\
NGC~5746& 0.05-0.15 & $<3.0 \times 10^{39}$ &  $<1.3 \times 10^{40}$  & $<5.9\times10^{9}$ & $<4.0\times10^{-5}$ & $>5.7$ & $<1.0$\\
ESO~445-081&0.05-0.15&$<2.9 \times10^{40}$&  $<1.4 \times 10^{41}$  & $<4.1\times10^{10}$& $<6.0\times10^{-4}$ & $>3.8$ & $<10.9$\\
STACK$^{\rm \dagger}$& 0.05-0.15 & $<2.0\times 10^{39}$  &  $<7.3 \times 10^{39}$  & $<3.2\times10^{9}$ &$<3.8\times10^{-4}$ & $>6.0$ & $<0.5$ \\
\hline 
NGC~266  & 0.15-0.30 & $<4.0 \times 10^{40}$ &  $<1.9 \times 10^{41}$  & $<7.3\times10^{10}$ & $<4.4\times10^{-4}$ &$>5.3$ & $<13.8$\\
NGC~1097& 0.15-0.30 & $<2.9 \times 10^{39}$ &  $<1.2 \times 10^{40}$  & $<8.4\times10^{9}$ &$<2.6\times10^{-4}$ & $>8.9$ & $<0.9$\\
NGC~2841& 0.15-0.30 & $<1.7 \times 10^{39}$ &  $<6.6 \times 10^{39}$  & $<1.2\times10^{10}$ &$<1.0\times10^{-4}$ & $>22.4$ & $<0.5$\\
NGC~5005& 0.15-0.30 & $<5.0 \times 10^{39}$ &  $<1.9 \times 10^{40}$  & $<1.3\times10^{10}$ & $<2.7\times10^{-4}$ & $>8.6$ & $<1.5$\\
NGC~5170&0.15-0.30 &  $<4.3 \times 10^{39}$ &  $<2.2 \times 10^{40}$  & $<1.4\times10^{10}$ & $<3.0\times10^{-4}$ & $>7.8$ & $<1.8$\\
NGC~5529& 0.15-0.30 & $<8.0 \times 10^{39}$ &  $<3.0 \times 10^{40}$  & $<2.1\times10^{10}$ &$<2.7\times10^{-4}$ & $>8.4$ & $<2.4$\\
NGC~5746& 0.15-0.30 & $<2.1 \times 10^{39}$ &  $<9.0 \times 10^{39}$  & $<1.3\times10^{10}$ & $<1.2\times10^{-4}$ & $>18.3$ & $<0.7$\\
ESO~445-081&0.15-0.30&$<6.4 \times10^{40}$&  $<3.1 \times 10^{41}$  & $<1.7\times10^{11}$ & $<3.3\times10^{-4}$ & $>6.8$ & $<24.4$\\
STACK$^{\rm \dagger}$& 0.15-0.30 &  $<2.7 \times 10^{39}$ &  $<9.8 \times 10^{39}$   & $<1.0\times10^{10}$ &$<1.6\times10^{-4}$ & $>14.4$ & $<0.7$\\
\hline \\
\end{tabular} 
\end{minipage}
$^{\dagger}$ Based on the stacked data of a sub-sample of six galaxies: NGC~1097, NGC~2841, NGC~5005, NGC~5170, NGC~5529, and NGC~5746.   \\
\label{tab:values}
\end{table*}

\section{Data reduction}
\subsection{\textit{Chandra} X-ray observations}
\label{sec:chandra}
The analysis of the \textit{Chandra} data was performed with standard \textsc{ciao} tools (version 4.6 and \textsc{caldb} version 4.6.2). For each galaxy we analyzed all available ACIS-S or ACIS-I observations. The total combined exposure time of the analyzed data is  $648.5$ ks. The analyzed \textit{Chandra} observations are listed in Table \ref{tab:list2}. 	

The first step of the data analysis was to reprocess all observations using the \textsc{chandra\_repro} task, which results in enhanced data quality and better calibrated observations. After this we identified and removed time intervals that were contaminated with  background flares. Since the detectors are most sensitive to flares in the $2.3-7.3$ keV band, we used this energy range to identify high background periods. For each observation, we excluded those time intervals that were $2\sigma$ above the mean count rate level. Removing the flares resulted in a $\approx12\%$ drop in the exposure times, and a total net exposure time of $574.2$ ks. However, note that the ACIS-S detectors are more susceptible to background flares, hence a larger fraction of contaminated time intervals were excluded from these data (Table \ref{tab:list2}). 

In general, bright point sources can add a significant contribution to the truly diffuse gaseous emission, hence it is necessary to exclude them when studying diffuse X-ray emitting components. To detect bright point sources, we ran the \textsc{wavdetect} task using the parameters described in \citet{bogdan08}. This procedure results in sufficiently large source cells that include $\gtrsim97\%$ of the source counts. These source cells were used to mask out bright point sources from the further study of the diffuse emission. 

To account for vignetting effects and convert the observed counts to flux units, we produced exposure maps assuming a thermal plasma model (\textsc{apec} in \textsc{Xspec}). The temperature of the model was $0.2$ keV and the metal abundances were set to 0.1 Solar \citep{grevesse98}. The particular choice of spectrum was motivated by the results of our earlier observations \citep{bogdan13a} and the Illustris Simulation. Indeed, Illustris predicts similar gas temperatures for a hot corona surrounding a galaxy with $M_{\rm \star}=10^{11} \ \rm{M_{\odot}}$, which is close to the median stellar mass of our sample. Additionally, we observed sub-solar metal abundances ($\sim0.1$ Solar) for NGC~1961 and NGC~6753, which is in fair agreement with that predicted by Illustris (Section \ref{sec:temp}). 

Since we aim to study the faint diffuse X-ray emission, it is crucial to precisely account for the background emission. We used two different approaches to subtract the background components. For the more distant galaxies in our sample, we used local fields to subtract the background level. Using internal background fields offers a robust way to subtract all background components without the need to renormalize the background level. We note that due to the large virial radii of the galaxies in our sample, the applied background regions may be within the extended coronae of the galaxies. However, at large radii  ($\gtrsim0.3r_{200}$) the gas is expected to have very low density (see Figure \ref{fig:density}), implying that the extended coronae remain well below the detection threshold. To confirm this picture, we extracted surface brightness profiles of the sample galaxies in the $0.3-2$ keV energy range, and found that the vignetting corrected profiles do not show variations at these large radii. This hints that at radii $\gtrsim0.3r_{200}$ the contribution of gaseous coronae are negligible compared to the sky and instrumental background components. For the more nearby galaxies in our sample, we relied on the ACIS blank-sky background files to subtract the background components. To ensure that soft X-ray sky background of the sample galaxies is similar to that of the background maps, we checked the fluxes of ROSAT All-Sky Survey R4-R5 band images at the position of the galaxies \citep{snowden97}. We found that the galaxies in our sample have similar soft X-ray brightness to those used for constructing the ACIS background files. Due to the stable spectrum of the instrumental background, the \textit{Chandra} blank-sky files can be tailored to a specific observation. To this end, we normalized the blank-sky background files using the observed count rate ratios in the $10-12$ keV band. The accuracy of these background subtraction methods were compared for those galaxies, where both procedures could be applied. We concluded that the background subtracted data in these galaxies are in good agreement with each other, and the accuracy of background subtraction is a few per cent.

\subsection{Near- and far-infrared data}
\label{sec:fir}
The near-infrared images offer a robust means to trace the stellar light. Therefore, we rely on the K-band images of the Two-Micron All Sky Survey (2MASS) to derive the stellar mass of the galaxies \citep{jarrett03}. Using the K-band magnitudes and the distances of the galaxies we derive the absolute K-band magnitudes and luminosities following  the procedure described in  \citet{bogdan13a}. The K-band luminosities are converted to stellar mass using the mass-to-light ratios computed from the corresponding $B-V$ color indices and results of galaxy evolution modeling \citep{bell03}. 

The star-formation rates of the sample galaxies are computed from the 60 $\mu$m and 100 $\mu$m far-infrared fluxes obtained by the Infrared Astronomical Satellite (\textit{IRAS}) and the relation established by \citet{kennicutt98}. To derive the total far-infrared luminosity of the sample galaxies, and hence derive the star-formation rates, we follow the methodology described in \citet{bogdan13a}. We conclude that all of the sample galaxies have low star-formation rates relative to their stellar mass, implying that none of the galaxies is a starburst system. The stellar masses  and star-formation rates of the sample galaxies are given in Table \ref{tab:list1}. 

\begin{figure}
  \begin{center}
    \leavevmode
      \epsfxsize=8.7cm\epsfbox{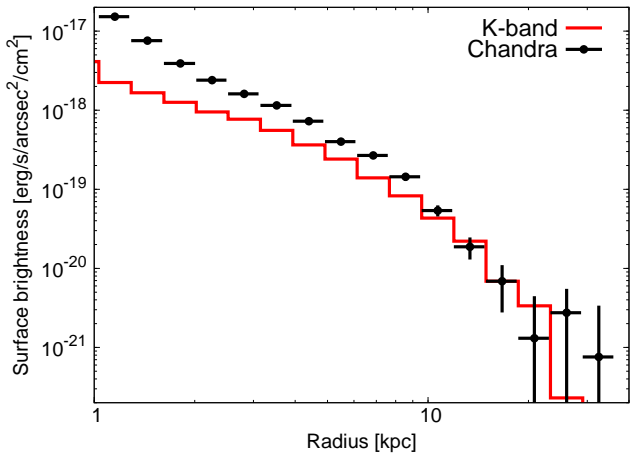}
      \caption{Stacked $0.3-2$ keV band X-ray surface brightness profile of the diffuse emission for the eight galaxies in our sample. To construct the profiles, we used circular annuli centered on the optical centroid of each galaxy. Vignetting correction is applied and all background components are subtracted.  Over plotted is the 2MASS K-band surface brightness profile, which traces the stellar light distribution. The K-band light is normalized to match the last X-ray detected bin at $\approx17$ kpc. Note that the X-ray surface brightness for the last three bins are consistent with 0.}
\vspace{0.5cm}
     \label{fig:profile}
  \end{center}
\end{figure}

\section{Results}
\subsection{Searching for hot X-ray coronae}
\label{sec:optical}
To identify hot coronae around the sample galaxies, we employed four different methods. First, we searched for diffuse X-ray emission by visually inspecting the background subtracted and vignetting corrected \textit{Chandra} images in a soft ($0.3-1$ keV), a broad ($0.3-2$ keV), and a hard band ($1-2$ keV). Probing different energy ranges is useful, as it offers the opportunity to detect an ionized gas component with relatively low ($<0.3$ keV) or higher ($>0.3$ keV) gas temperatures. While a diffuse X-ray glow is associated with some of the spiral galaxies in our sample, this emission is focused in the innermost regions. However, within these regions several additional X-ray emitting components are present, such as unresolved low-mass and high-mass X-ray binaries, cataclysmic variables, active binaries, young stars, young stellar objects, and diffuse gas associated with star formation   \citep{gilfanov04,revnivtsev08,bogdan11,mineo12}. Given the overall difficulty to assess the importance of each of the above listed components to the diffuse gaseous emission, we must explore the hot coronae beyond the optical radii of the galaxies. 

\begin{figure*}[t]
  \begin{center}
    \leavevmode
    \vspace{0.5cm}
      \epsfxsize=8.7cm\epsfbox{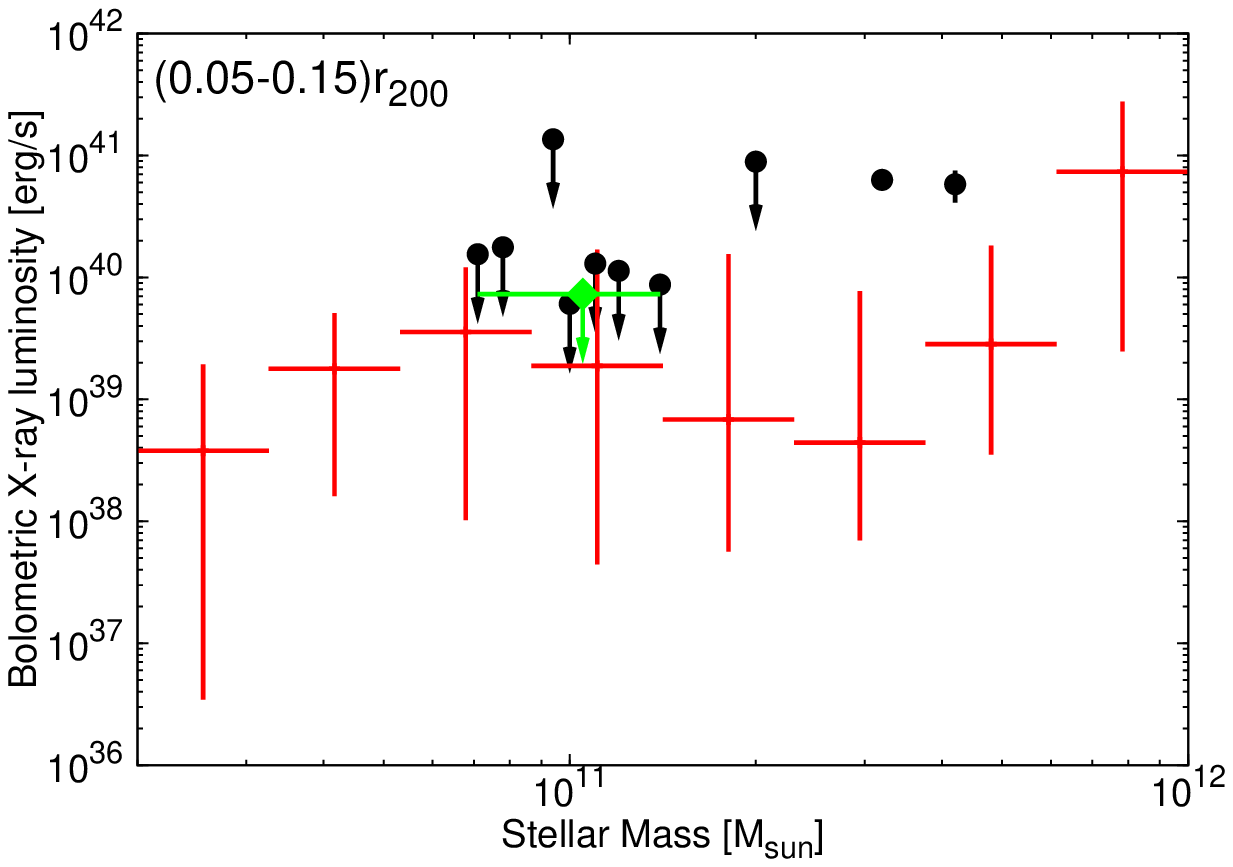}
      \epsfxsize=8.7cm\epsfbox{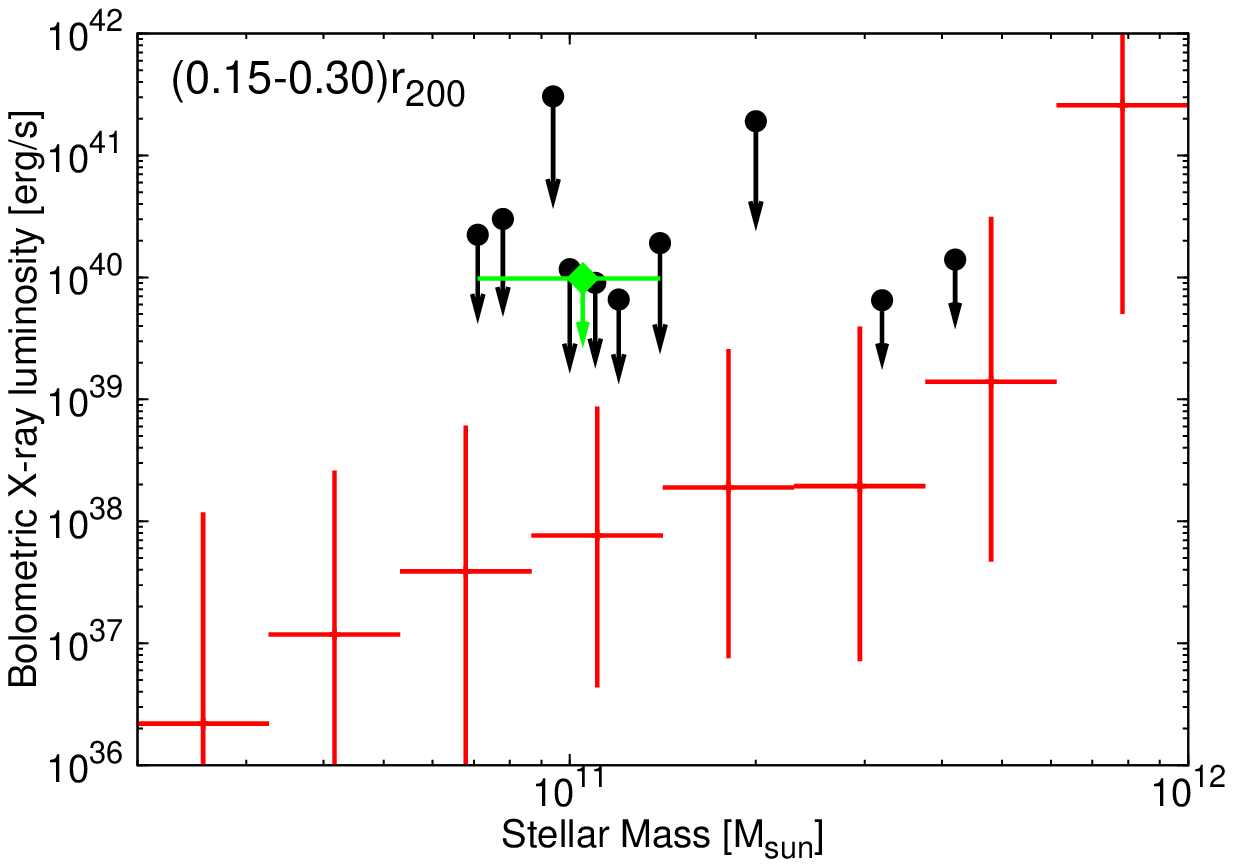}
\vspace{0.5cm}
      \epsfxsize=8.7cm\epsfbox{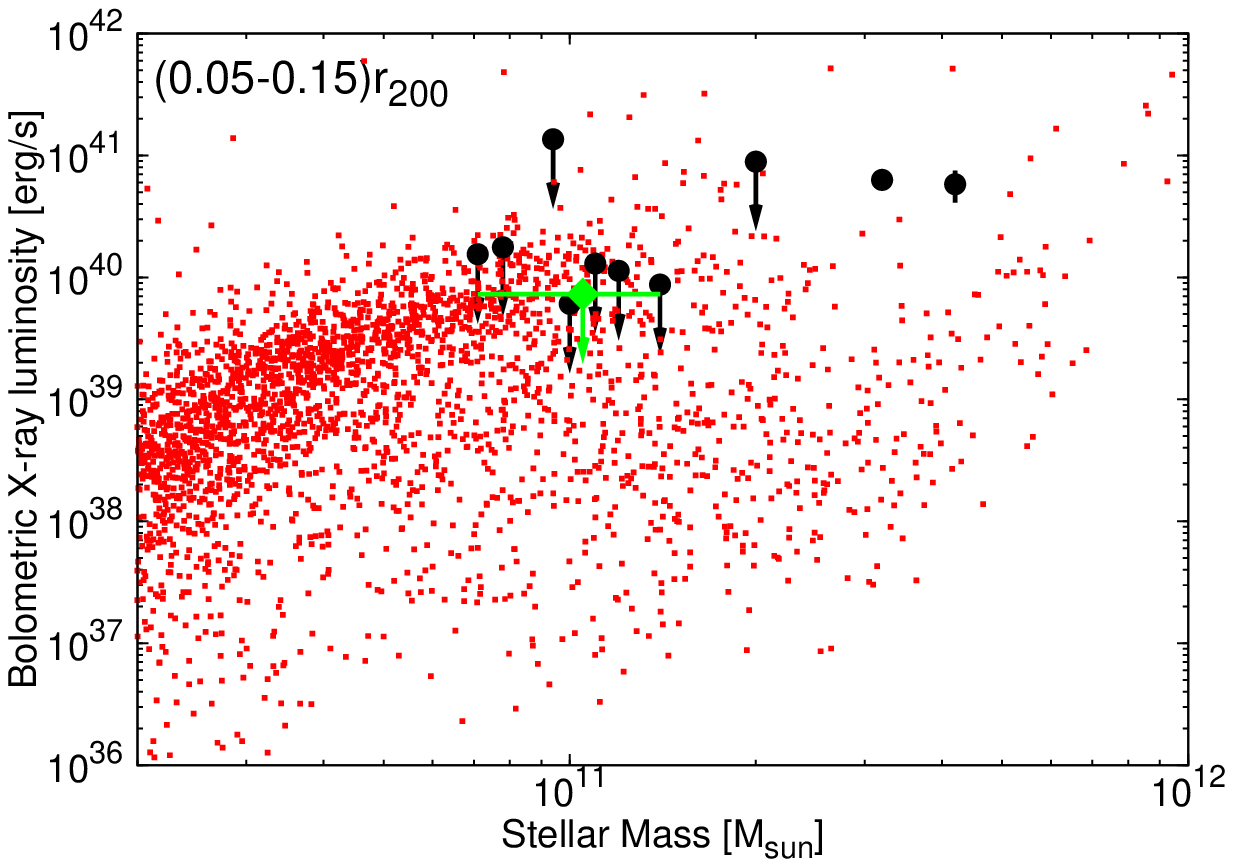}
      \epsfxsize=8.7cm\epsfbox{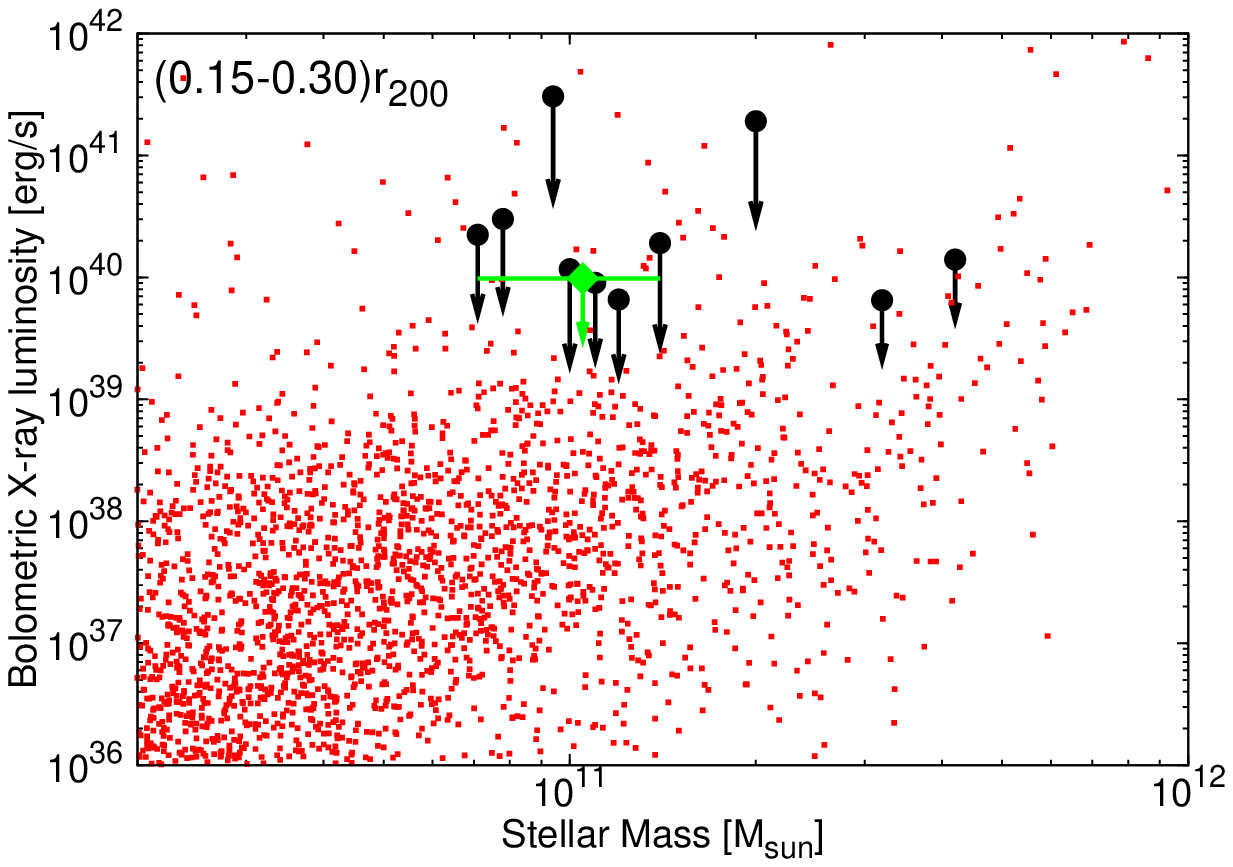}
       \epsfxsize=8.7cm\epsfbox{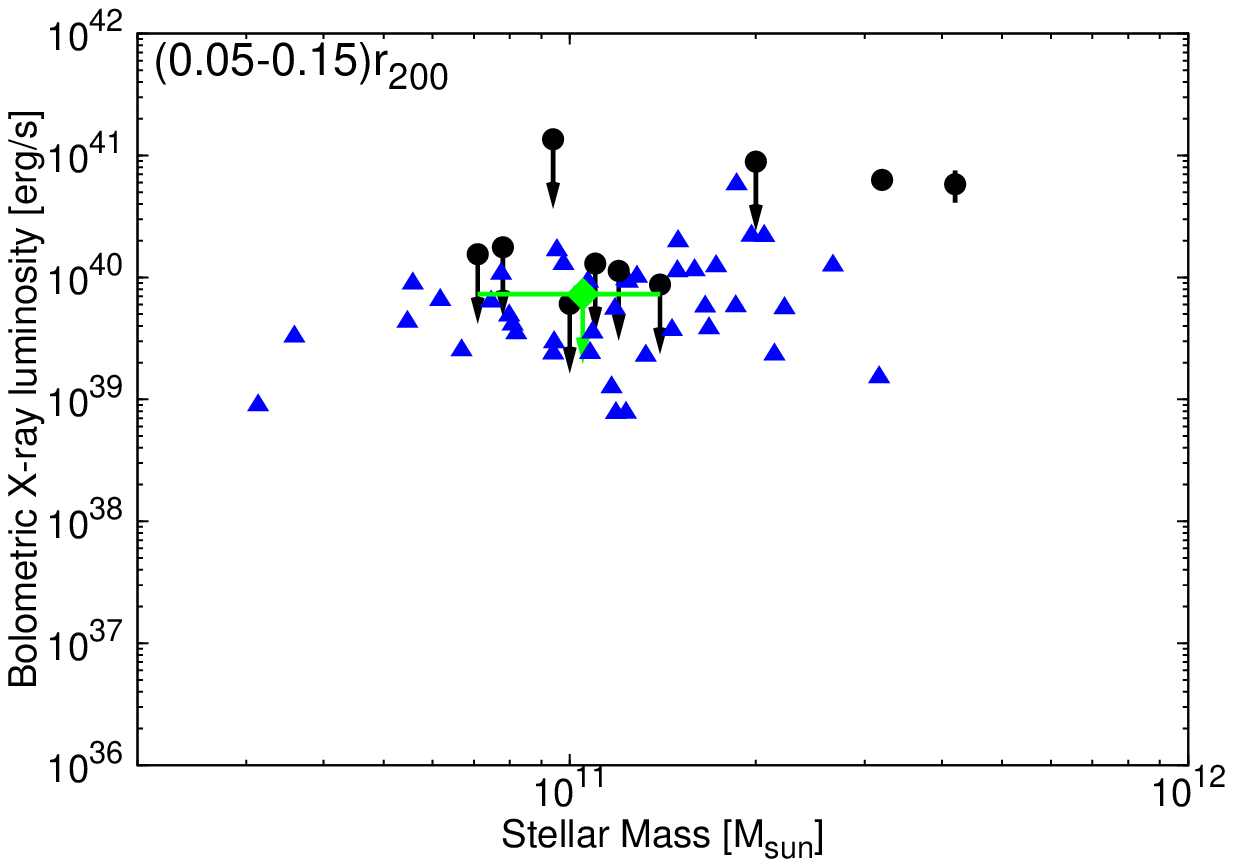}
      \epsfxsize=8.7cm\epsfbox{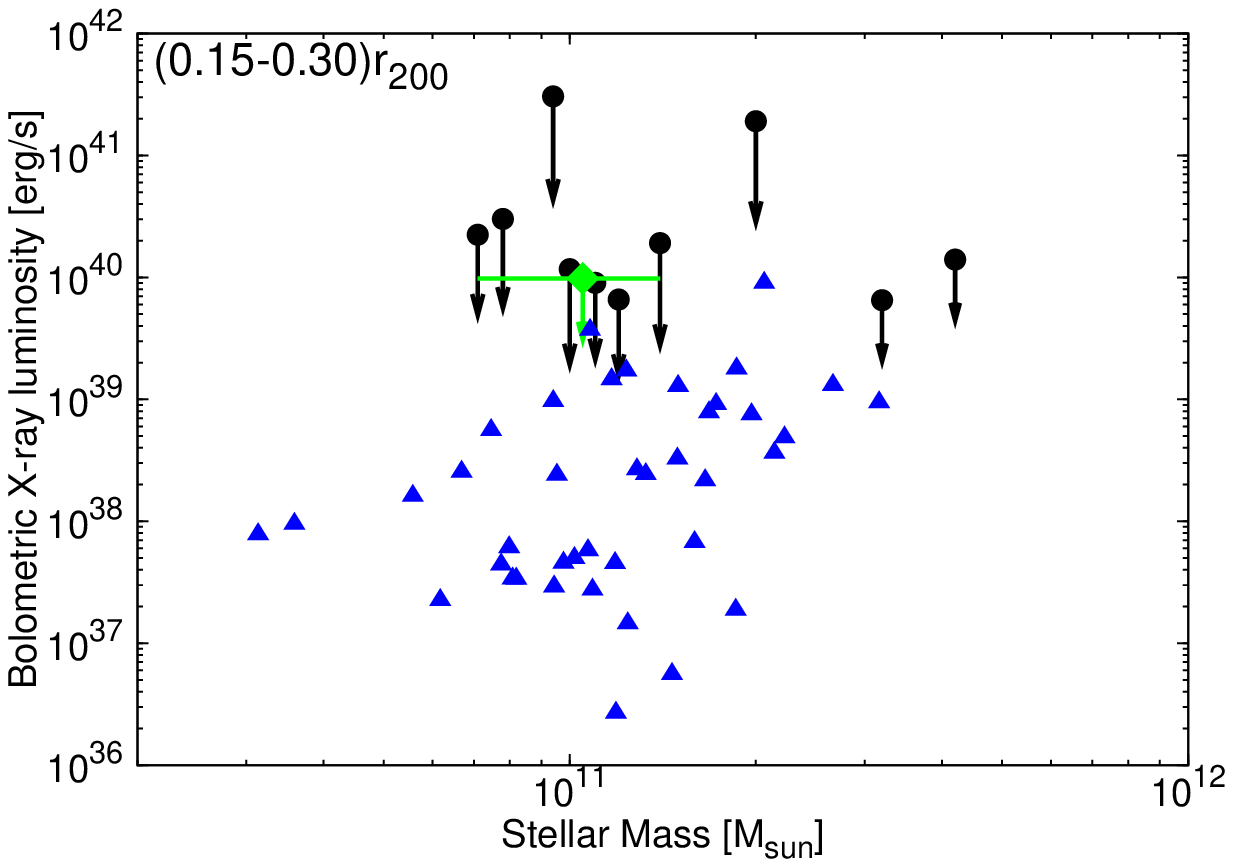}
      \caption{Bolometric X-ray luminosity in the $(0.05-0.15)r_{200}$ (left panels) and in the $(0.15-0.30)r_{200}$ (right panels) region as a function of the stellar mass.  The upper panels show the $10-90$ percentile range of the predicted X-ray luminosities in the Illustris Simulation for all simulated galaxies within the stellar mass range of $(2\times10^{10} - 10^{12} \ \rm{M_{\odot}}$. For each stellar mass bin the median X-ray luminosity is marked with the horizontal bars. The small points in the middle panels depict the individual predicted X-ray luminosities of all simulated galaxies. In the lower panel, triangles show the simulated X-ray luminosity of 42 massive spiral galaxies from Illustris.  The large filled circles points show the observed X-ray luminosities for NGC~1961 and NGC~6753  \citep{bogdan13a}, along with the $3\sigma$ upper limits presented in this work. The large filled diamonds with horizontal bars represent the $3\sigma$ upper limits on the X-ray luminosity derived from the stacked data of a sub-sample of the six most nearby galaxies. The uncertainty in the observed X-ray luminosity  for NGC~6753 is consistent with the size of the symbol \citep{bogdan13a}. Note that there is a large scatter in the predicted X-ray luminosity at stellar masses $>10^{11} \ \rm{M_{\odot}}$. The scatter and the nearly constant predicted X-ray luminosity for galaxies with stellar mass above $\sim10^{11} \ \rm{M_{\odot}}$ is the consequence of powerful radio-mode AGN feedback, which ejects a significant amount of hot gas from the halos.}
\vspace{0.5cm}
     \label{fig:lx}
  \end{center}
\end{figure*}

\begin{figure*}[t]
  \begin{center}
    \leavevmode
    \vspace{0.5cm}
      \epsfxsize=8.7cm\epsfbox{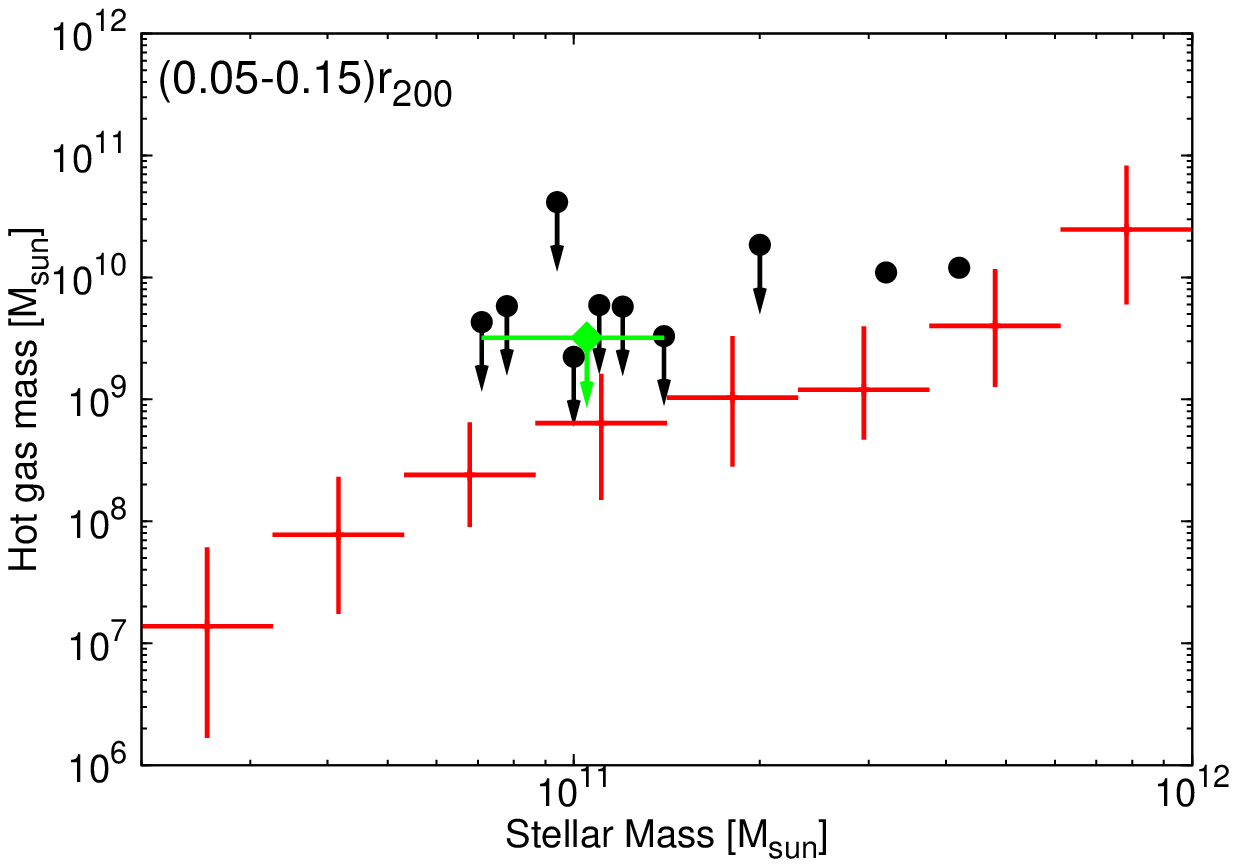}
      \epsfxsize=8.7cm\epsfbox{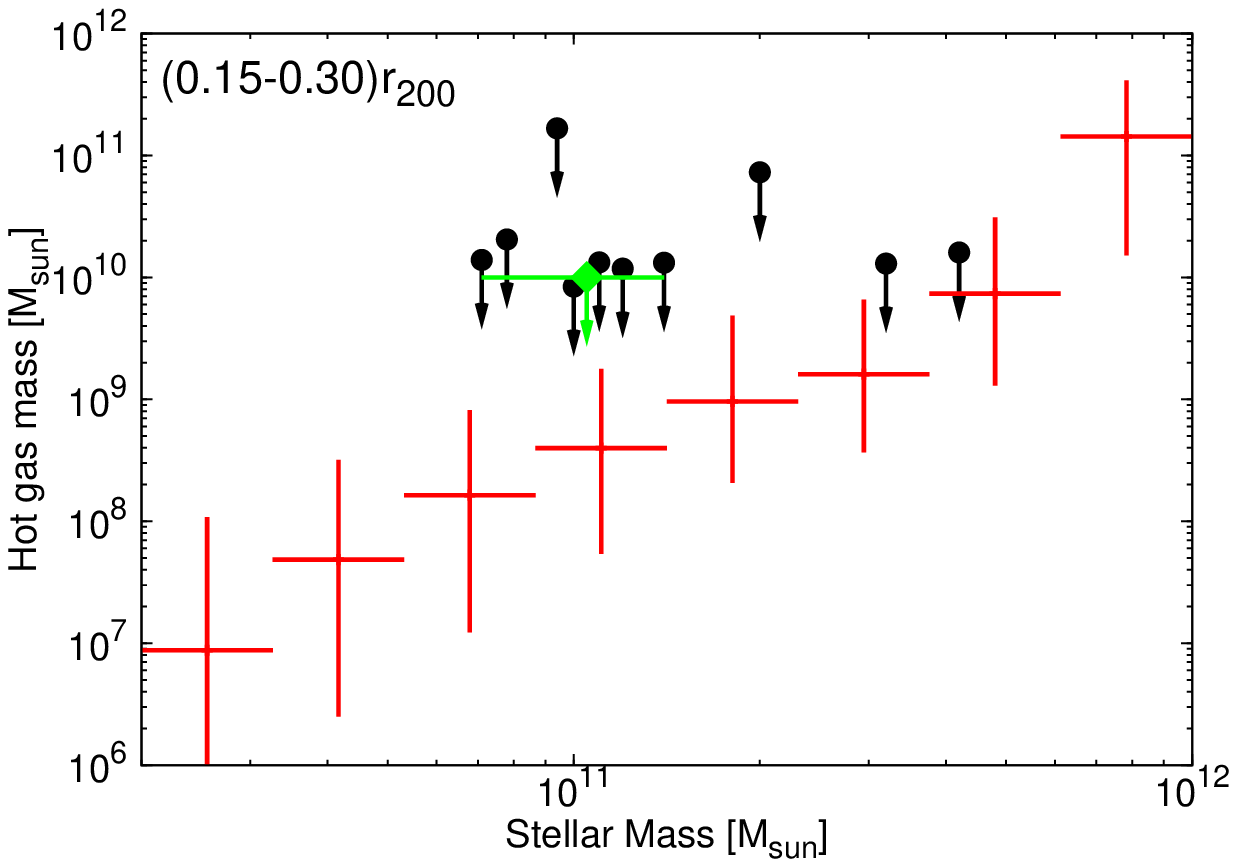}
      \epsfxsize=8.7cm\epsfbox{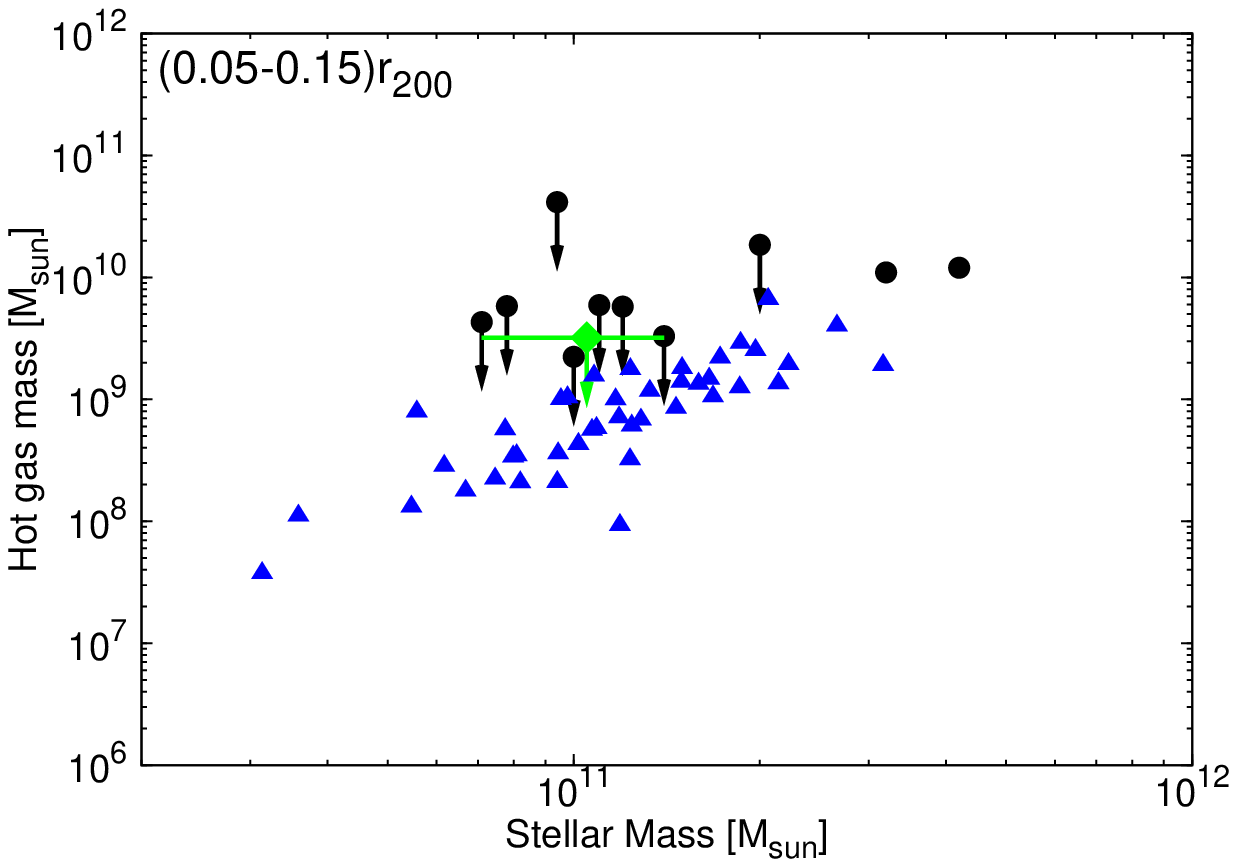}
      \epsfxsize=8.7cm\epsfbox{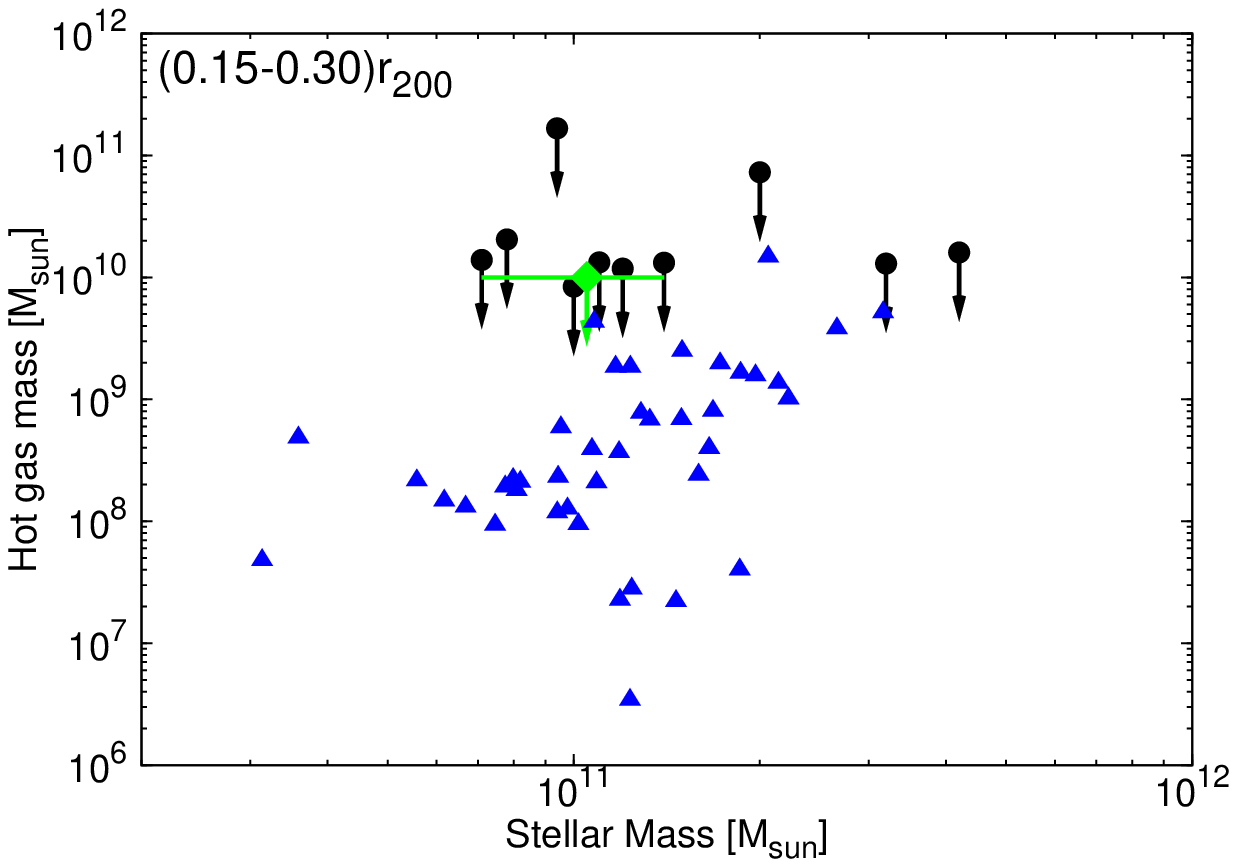}
      \caption{Hot X-ray gas mass confined within the $(0.05-0.15)r_{200}$ (left panel) and in the $(0.15-0.30)r_{200}$ (right panel) regions as the function of the stellar mass. In the upper panels, the $10-90$ percentile range of the simulated gas masses are shown with vertical bars, while the horizontal bars show the median gas mass at the given stellar mass range. In the lower panels, we depict the simulated gas masses of the 42 textbook spiral galaxies. Note that only those gas particles were included in the simulations, which have a temperature of at least $0.1$ keV. Large filled circles represent the observed $3\sigma$ upper limits and the observed hot gas masses for NGC~1961 and NGC~6753 \citep{bogdan13a}. The uncertainty in the observed gas mass for NGC~1961 and NGC~6753 is consistent with the size of the symbols \citep{bogdan13a}. The large filled diamonds with horizontal bars represent the $3\sigma$ upper limits on the X-ray luminosity derived from the stacked data of a sub-sample of the six most nearby galaxies.}
\vspace{0.5cm}
     \label{fig:mass}
  \end{center}
\end{figure*}

As a second approach, we probed the existence of hot coronae by constructing radial surface brightness profiles for each galaxy in the $0.3-1$ keV, $0.3-2$ keV,  and $1-2$ keV bands. To construct the profiles we utilized the background subtracted and vignetting corrected data, and extracted circular annuli centered on the optical centroid of each galaxy. The X-ray profiles are compared with the stellar light distribution inferred from the 2MASS K-band images. Based on these profiles we did not detect an extended X-ray emitting component beyond the optical radii. 

Third, to increased the signal-to-noise ratio of the data, we co-added (i.e. stacked) the  individual surface brightness distributions in the three studied energy ranges. In Fig. \ref{fig:profile}, we show the stacked $0.3-2$ keV band surface brightness profile of the diffuse X-ray emission as well as the stacked K-band light distribution. While the X-ray and K-band surface brightness distributions are markedly different, the stacked profile does not reveal a statistically significant diffuse X-ray emitting component beyond $\sim20$ kpc. In agreement with these, the soft and hard band profiles also did not reveal the presence of diffuse X-ray emission beyond the extent of the stellar light. 

Finally, we probed the existence of an extended X-ray emitting component in the $(0.05-0.15)r_{\rm 200}$ and the $(0.15-0.30)r_{\rm 200}$ regions in the soft, broad, and hard band \textit{Chandra} data. However, we could not identify statistically significant diffuse X-ray emission in either radial bin. As before, we increased the signal-to-noise ratios by stacking the X-ray photons in the three energy ranges. The stacked data also did not reveal a statistically significant signal. Interestingly, for NGC~266 we detect a moderately significant ($\sim2\sigma$) X-ray emitting component in the $(0.05-0.15)r_{\rm 200}$ region, which possibly originates from its hot gaseous corona. We detect $137 \pm 60$ net counts in the $0.3-2$ keV energy range, where the error range refers to the statistical uncertainties. Given the low signal-to-noise ratio, the importance of  systematic uncertainties  associated with the background subtraction is non-negligible. Specifically, depending on the background subtraction method, the obtained signal exhibits somewhat different ($\sim1.5\sigma-2.5\sigma$) statistical significance. Despite the low significance detection of the diffuse emission around NGC~266, we note that the observed count rates are broadly consistent with our earlier results presented in \citet{bogdan13b} -- for further discussion see Section \ref{sec:disc}. 

These evidences, both individually and combined, point out that statistically significant diffuse X-ray emission is not detected beyond the optical radii ($\gtrsim 20$ kpc) of the sample galaxies.

 \begin{figure*}[t]
  \begin{center}
    \leavevmode
      \epsfxsize=8.7cm\epsfbox{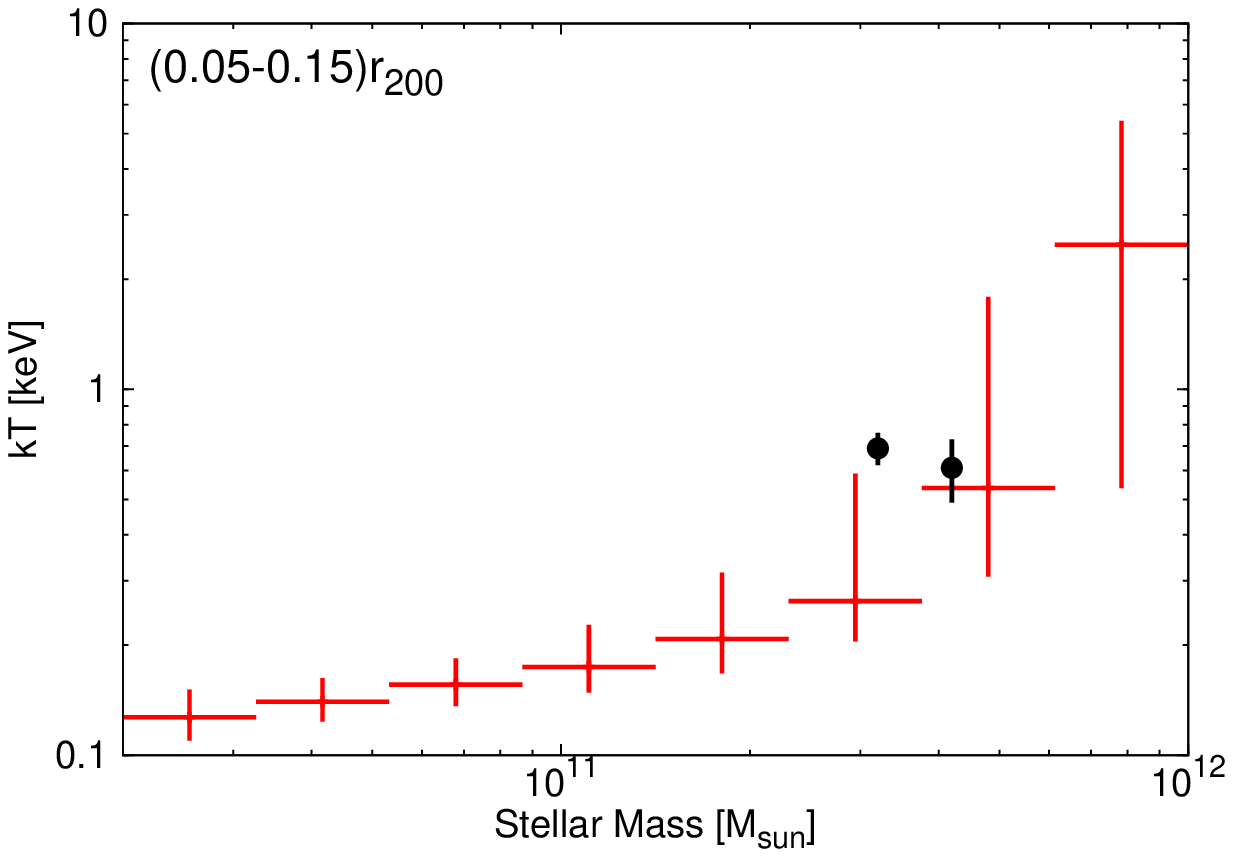}
      \epsfxsize=8.7cm\epsfbox{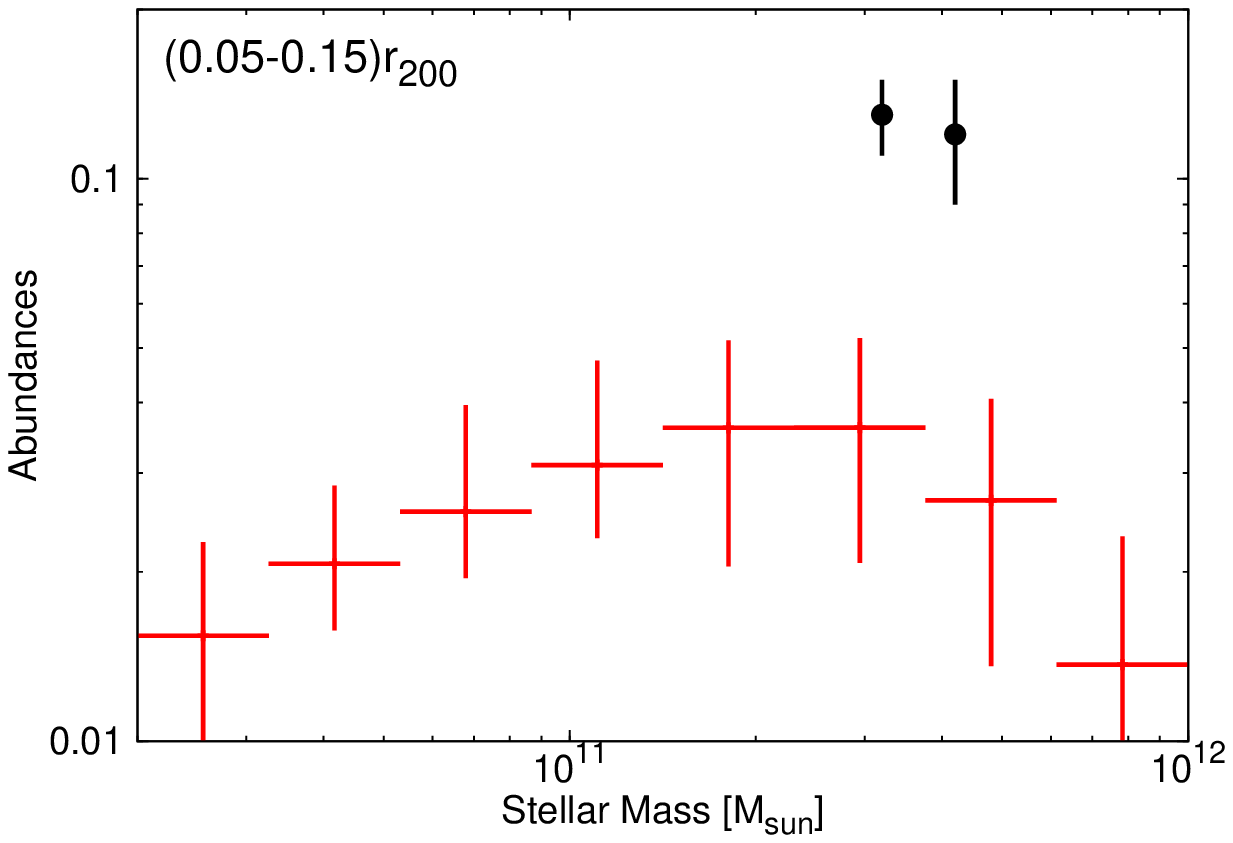}
      \caption{The left and right panels shows the temperature and metal abundances of the hot gas  within the $(0.05-0.15)r_{200}$ region as a function of the stellar mass, respectively. Both the gas temperature and the abundance refers to luminosity-weighted quantities. The vertical bars show the $10-90$ percentile range, and the horizontal bars show the median values for each stellar mass bin. The observed values for NGC~1961 and NGC~6753 are shown as a comparison. Note that to date only these two galaxies offer a measurement of the gas temperature and abundance beyond their optical body.}
\vspace{0.5cm}
     \label{fig:temp}
  \end{center}
\end{figure*}

\subsection{Constraining the gas parameters} 
In the absence of statistically significant detections of the extended emission beyond the optical radii, we compute $3\sigma$ limits on the characteristics of the hot X-ray coronae. To this end, we derive the $3\sigma$ upper limits on the X-ray counts in circular annuli with radii of  $(0.05-0.15)r_{\rm 200}$ and $(0.15-0.30)r_{\rm 200}$ following \citet{bogdan13a}. We utilize all available \textit{Chandra} detectors to maximize the number of detected counts. To convert the counts to physical units, we assume an optically-thin thermal plasma emission (\textsc{apec}) model with $kT=0.2$ keV temperature, 0.1 Solar abundance, and Galactic column density. To account for the vignetting effects and the different sensitivity of various detectors, we employ the tailor-made response files for each observation. The  upper limits on the X-ray counts were converted to the normalization of the \textsc{apec} model using:
$$ N= \frac{10^{-14}}{4 \pi D^2} \int n_e n_H dV \ ,$$ 
where $D$ is the distance of the galaxy, while $n_e$ and $n_H$ are the electron and hydrogen number densities, respectively. Based on the upper limits on the emission measure ($\int n_e n_H dV$), we compute upper limits on the X-ray luminosity, average gas density, and confined gas mass within the volume that corresponds to the radial ranges of $(0.05-0.15)r_{\rm 200}$ and $(0.15-0.30)r_{\rm 200}$. Note that in these calculations we assumed that the hot gas has a spherically symmetric distribution and constant gas density within the given bins. Additionally, we derive lower limits on the cooling time, which is computed as $t_{\rm{cool}}=(3kT)/(n_{\rm{e}} \Lambda(T))$, where $\Lambda(T)$ is the cooling function. Finally, we also derive upper limits on the cooling rates of the gas as $\dot M = M_{gas}/t_{\rm cool}$, which allows us to probe whether cooling of the hot gaseous coronae can balance the ongoing star-formation in the disk (Section \ref{sec:disc}).

To place more stringent constraints on the properties of hot coronae, we stacked the X-ray photons along with the corresponding exposure maps for the six most nearby galaxies in our sample, that is for NGC~1097, NGC~2841, NGC~5005, NGC~5170, NGC~5529, and NGC~5746. The stacked data allow us to compute the average $3\sigma$ limits on the gas parameters for spiral galaxies in the stellar mass range of $(0.7-1.4)\times10^{11} \ \rm{M_{\odot}}$. Thanks to the increased signal-to-noise ratios, these limits are more stringent than those obtained for individual galaxies. To derive the $3\sigma$ limits on the gas characteristics for the stacked data, we utilized the median distance, virial radius, and column density of the sub-sample, and followed the procedure described for individual galaxies. The limits are given in Table \ref{tab:values}. 

Although the main focus of this work is to probe the characteristics of hot coronae around lower mass spirals, we extend the present sample by our earlier detections of the coronae around NGC~1961 and NGC~6753 \citep{bogdan13a}. Including these galaxies allows us to confront the Illustris Simulation with observational results in the stellar mass range of $(0.7-4.2)\times10^{11} \ \rm{M_{\odot}}$.

\section{Comparison with Illustris}
\label{sec:simulation}
\subsection{The Illustris Simulations}
The Illustris Project (http://www.illustris-project.org) comprises a series of cosmological hydrodynamic simulation runs in a $(106.5 \ \rm{Mpc})^3$ volume performed with the moving mesh code \textsc{arepo} \citep{springel10}. The simulations were carried out at a series of resolutions, of which the highest resolution, Illustris-1 (or simply Illustris), has a mass resolution of $m_{\rm{DM}}=6.26 \times 10^6 \rm{M_{\odot}}$ and $m_{\rm{baryon}} \backsimeq 1.26 \times 10^6 \rm{M_{\odot}}$ for the dark matter and baryonic components, respectively. At $z=0$ the co-moving plummer equivalent gravitational softening lengths are 1.4 kpc for the dark matter and 0.7 kpc for baryonic collisionless particles, while for gas it is adaptively set by the cell size but cannot be lower than 0.7 kpc. In this paper, we rely on the results of the Illustris run. The Illustris Simulation discussed here adopts the cosmological parameters that are consistent with the WMAP-9 measurements \citep{hinshaw13}.  

Illustris incorporates the crucial physical processes that are indispensable to describe galaxy formation and evolution\citep{vogelsberger14a,vogelsberger14b,genel14}. Specifically, it includes primordial and metal line cooling, describes the stellar evolution and the corresponding feedback processes, traces the chemical enrichment processes by modeling nine elements, seeds supermassive black holes and follows their evolution through accretion and mergers, and models the energetic quasar-mode and radio-mode feedback of supermassive black holes. Details of the incorporated physics modules were discussed in previous works \citep{vogelsberger13,torrey14}. A major consequence of the implemented stellar and AGN feedback  is that the Illustris simulation can produce a stellar mass -- halo mass relation, which is in good agreement with that observed from abundance matching studies. Additionally, Illustris can successfully reproduce the observed cosmic star-formation density, and can result in a realistic galaxy population. Overall,  Illustris offers an ideal framework to confront the observed and simulated properties of hot X-ray coronae.

Following our observational approach, we derive the characteristics of the hot gas in Illustris in two radial ranges. For the inner and outer regions, we utilize the $0.05r_{\rm 200}<r<0.15r_{\rm 200}$ and $0.15r_{\rm 200}<r<0.30r_{\rm 200}$ regions, respectively. Since we aim to study the hot gas, we only included gas particles, which have a temperature of at least $1.16\times10^6$ K. To derive the X-ray luminosity and other properties of the gas we follow the procedure described in \citet{navarro95}.

\subsection{X-ray luminosities and gas masses}
Since the X-ray luminosity and gas mass of  hot coronae have been widely used to test galaxy formation models, we first confront these quantities with those predicted by Illustris. To facilitate the comparison with previous and subsequent works, we perform the comparison in the $(0.05-0.15)r_{200}$ and in the $(0.15-0.30)r_{200}$ radial ranges. 
 
\begin{figure*}[t]
  \begin{center}
    \leavevmode
      \epsfxsize=8.7cm\epsfbox{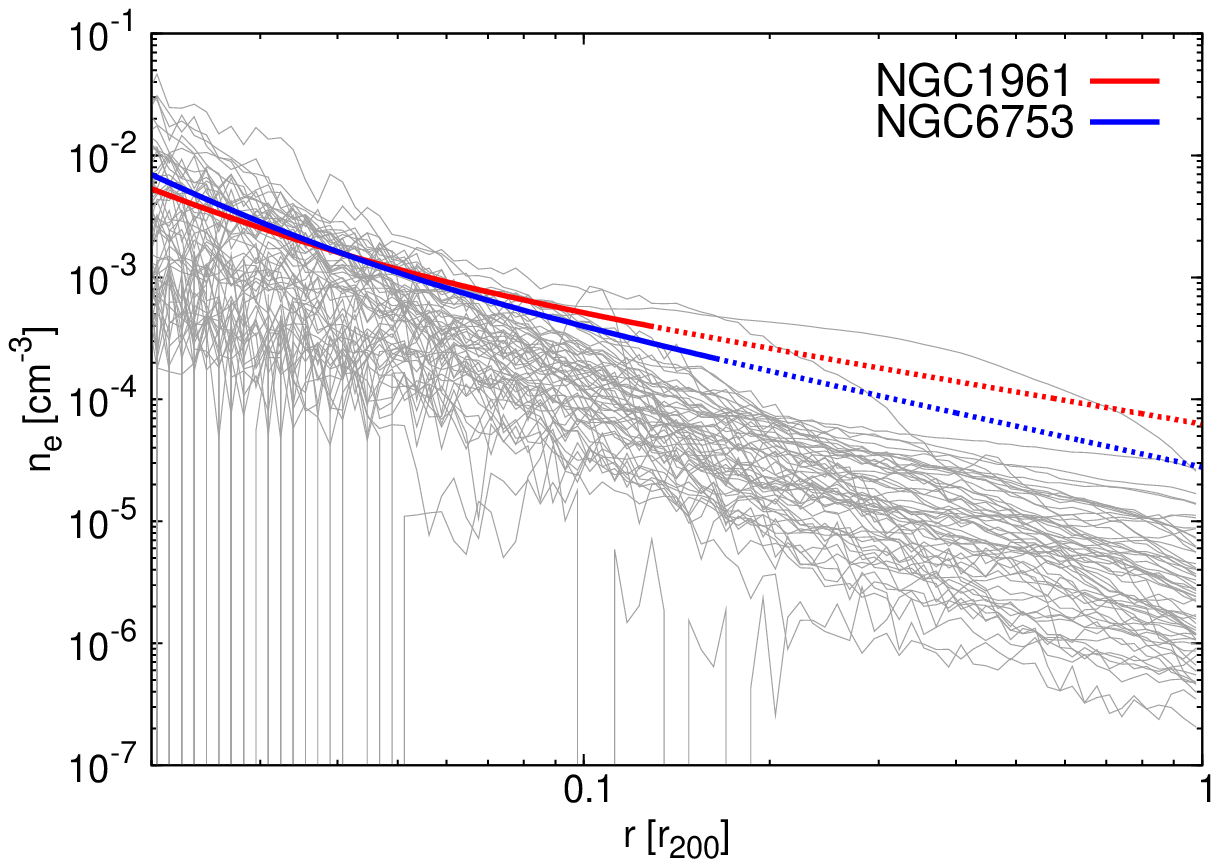}
      \epsfxsize=8.7cm\epsfbox{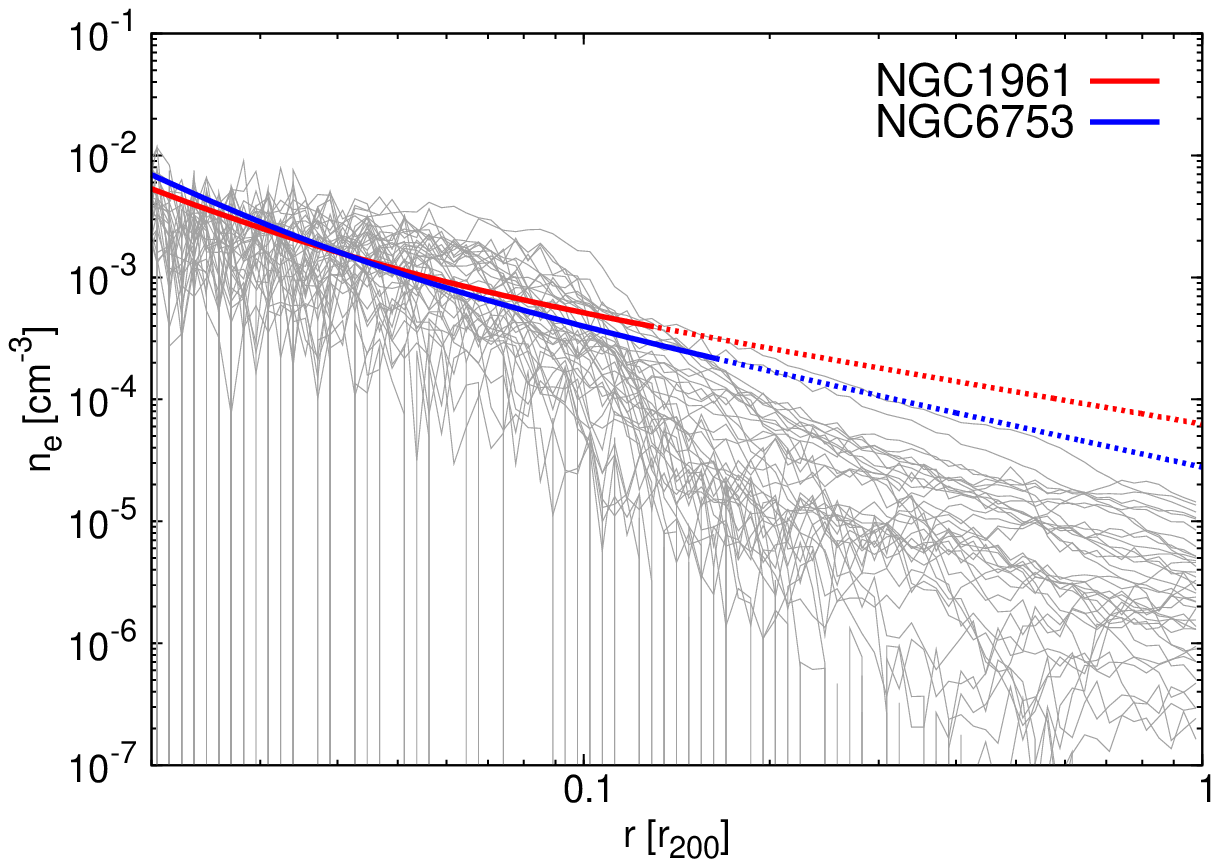}
      \caption{Electron density profiles as a function of radius. In the left panel, the light grey lines show the simulated profiles for the 61 galaxies in the Illustris simulation, whose stellar masses is in the range of $(3.0-4.5)\times10^{11} \ \rm{M_{\odot}}$. In the right panel, the light grey lines represent the 42 massive spiral galaxies. Note that the stellar mass of these textbook spirals is in the range of  $(0.3-3.2)\times10^{11} \ \rm{M_{\odot}}$, and hence remain below the stellar masses of NGC~1961 or NGC~6753. The thick red and blue lines represent the observed density profiles for NGC~1961 and NGC~6753, respectively. The short dashed red and blue lines show the extrapolation of the density profiles based on the best-fit modified $\beta$-models \citep{bogdan13b}. The regions $ \lesssim0.02r_{200}$ are not plotted, as they are not well resolved close to the gravitational softening scale, and they are not sampled by a large number of particles, resulting in larger uncertainties.  The radius is measured in units of  $r_{200}$.}
\vspace{0.5cm}
     \label{fig:density}
  \end{center}
\end{figure*}

When investigating the full sample, Illustris predicts large scatter in the X-ray luminosity and gas mass at every stellar mass (Figs. \ref{fig:lx}, \ref{fig:mass}). Moreover, the median X-ray luminosities and gas masses show an increasing trend in the stellar mass range of $(0.2-1)\times 10^{11}  \ \rm{M_{\odot}}$, while in $M_{\rm \star}=(1-6)\times10^{11} \ \rm{M_{\odot}}$ these values decline or remain broadly constant.  The reason for the relatively low luminosities and gas masses at $M_{\rm \star}\gtrsim10^{11} \ \rm{M_{\odot}}$ is that the radio-mode AGN feedback becomes dominant above these masses. As discussed in \citet{genel14}, the radio-mode feedback was tuned in a way that it can effectively suppress the star formation in massive halos, which in turn blows out a significant fraction of the gas from these massive halos, thereby resulting in less massive and less luminous X-ray coronae. For galaxies with less massive black holes, such as those residing in low-mass ellipticals or spirals, the radio-mode feedback is not dominant due to the weaker outbursts. This suggests that spiral galaxies in Illustris host more luminous X-ray coronae than ellipticals. To directly probe this possibility, we compare the median simulated X-ray luminosities for the ``textbook'' spiral and elliptical galaxies in the stellar mass range of  $(0.7-1.4)\times 10^{11}  \ \rm{M_{\odot}}$. Since the median stellar and total halo mass of both sub-samples are identical, the different X-ray luminosities cannot be attributed to differences in the depth of the gravitational potential well.  We find that within the $(0.05-0.15)r_{\rm 200}$ region the median luminosity for ``textbook'' spirals ($L_{\rm bol}=4.8\times10^{39} \ \rm{erg \ s^{-1}}$) exceeds the median luminosity of ``textbook'' ellipticals ($L_{\rm bol}=1.1\times10^{38} \ \rm{erg \ s^{-1}}$) by factor of $\sim44$.  This conclusion contradicts observational results. Indeed, due to their more luminous nature, gaseous coronae around massive ellipticals are routinely observed \citep{forman85,osullivan01,mathews03,bogdan11}, as opposed to the significantly fainter and less explored coronae around spirals. 

In the $(0.05-0.15)r_{\rm 200}$ region, the observed $3\sigma$ upper limits on the X-ray luminosities and gas masses are at the upper end of the simulated values when considering the full Illustris galaxy population. When focusing on the 42 simulated ``textbook'' spirals, we conclude that the observed  $3\sigma$ limits are at the upper end of the model predictions. Specifically, the observed limits on the luminosities and gas masses for the stacked sub-sample exceed the median simulated values for the ``textbook'' spirals. In the $(0.15-0.30)r_{\rm 200}$ range the median predicted luminosities and gas masses fall significantly short of the predicted upper limits. Based on these, we conclude that the Illustris predictions do not contradict the observational results for spiral galaxies in the stellar mass range of $(0.7-2.0)\times10^{11} \ \rm{M_{\odot}}$. However, the predicted luminosities and gas masses are significantly lower than those observed for NGC~1961 and NGC~6753, which can be attributed to the overly powerful radio-mode feedback present in Illustris.

\subsection{Gas temperature and abundance}
\label{sec:temp}
Although the non-detections of hot coronae do not allow us to constrain the gas temperatures and metal abundances, we utilize our previous measurements for NGC~1961 and NGC~6753 \citep{bogdan13a}. To this end, we confront the predicted luminosity-weighted temperatures and abundances with the observed values  within the $(0.05-0.15)r_{200}$ region. Since the 42 simulated ``textbook'' spirals are less massive than the observed galaxies, we focus on the entire Illustris population. 

While galaxies with  $M_{\rm \star}\lesssim10^{11} \ \rm{M_{\odot}}$ are expected to have  $0.1-0.2$ keV gas temperatures, the predicted temperatures shows a rapid increase for more massive galaxies. The observed gas temperatures for NGC~1961 and NGC~6753 are $kT\sim0.6-0.7$ keV, which agrees with those predicted by the Illustris simulation.  

The right panel of  Fig. \ref{fig:temp} reveals that Illustris predicts  $\sim0.01-0.05$ Solar median metal abundances for all galaxies in the studied stellar mass range. While the observed values exceed the simulated ones by a factor of a few, we do not consider this difference robust due to the notable systematic uncertainties associated with the abundance measurements \citep{bogdan13a}. To further constrain the major metal enrichment processes, deep X-ray observations are required, which can provide the sensitivity to account for the systematic uncertainties in the abundance measurements, thereby constraining the metal enrichment processes in greater detail.

\subsection{Electron density profiles}
To further probe Illustris, we confront the predicted electron density profiles with those observed for NGC~1961 and NGC~6753 \citep{bogdan13a}. The right panel of Fig. \ref{fig:density} depicts 61 Illustris galaxies with stellar and virial masses in the range of $(3.0-4.5)\times10^{11} \ \rm{M_{\odot}}$ and $(0.5-4.2)\times10^{13} \ \rm{M_{\odot}}$, respectively. These values are in agreement with those of NGC~1961 and NGC~6753. In the right panel of Fig. \ref{fig:density}, we compare the observed density profiles with the predicted density profiles of the 42 simulated ``textbook'' spiral galaxies. These galaxies, on average, are less massive than NGC~1961 and NGC~6753. 

When comparing the density profiles of the 61 massive simulated galaxies with those of NGC~1961 and NGC~6753, we conclude that the normalization and shapes of the profiles are broadly similar. However, the agreement is not perfect, because the average normalization of the simulated profiles are somewhat lower than observed for NGC~1961 and NGC~6753, and beyond $\sim0.1r_{\rm 200}$ the slope of the simulated profiles are steeper than the observed ones. As a caveat, we emphasize that beyond $\sim0.15r_{200}$ the density profiles for NGC~1961 and NGC~6753 are extrapolated, and hence carry significant uncertainties. Therefore, it is not clear how much weight should be given to the fact that the observed and simulated profiles exhibit somewhat different slopes at large radii.

A similar picture is obtained when we probe the electron density profiles of the 42  simulated``textbook'' spirals. Interestingly, these profiles have similar normalizations to those of NGC~1961 and NGC~6753, despite the fact that the Illustris galaxies have lower average stellar mass. Beyond $\sim0.1r_{200}$ the simulated profiles decline more steeply, which may be the combined effect of the lower gas mass in these halos and the powerful radio-mode AGN feedback, which ejects the hot gas to larger radii and/or removes it from the dark matter halos of galaxies.

\section{Discussion}
\label{sec:disc}
Since we are studying spiral galaxies with star-formation rates in the $\sim1-6 \ \rm{M_{\odot} \ yr^{-1}}$ range, it is important to probe if these galaxies could be prone to starburst driven winds at the present epoch. Indeed, theoretical calculations indicate and observational results confirm that in galaxies with the surface area specific supernova rate of $>40 \ \rm{SN \ Myr^{-1} \ kpc^{-2}}$ the detection of extraplanar emission is expected \citep{strickland04}. Note that the surface area is defined as $A=D_{25}^2$, where $D_{25}$ is the diameter of the isophote having the surface brightness of $25 \ \rm{mag \ arcsec^{-2}}$. To derive the specific supernova rates, we compute the supernova rate as $ R_{\rm{SN}} = 0.2 L_{\rm{FIR}}/10^{11} \ \rm{L_{\odot}}$, where $L_{\rm{FIR}}$ is the total far-infrared luminosity \citep{heckman90}, and use the $D_{25}$ radii given in Table \ref{tab:list1}. We find that the supernova rates are in the range of $0.005-0.068 \ \rm{yr^{-1}}$, and the observed area specific supernova rates are in the range of $\sim2-36 \ \rm{Myr^{-1} \ kpc^{-2}}$ for all galaxies but NGC~5005 and ESO~445-081.  This implies that we do not expect extraplanar gaseous emission driven by the energy input of supernovae for the six galaxies in our sample. While the specific supernova rates for  NGC~5005 and ESO~445-081 exceed the critical rate, they remain  below the values obtained for well-known starburst galaxies, such as M82 \citep{strickland04}. Therefore, while we may expect an outflow in NGC~5005 and ESO~445-081, these are likely less prominent than those in starburst galaxies, hinting that the hot gaseous coronae are not dominated by hot gas originating from starburst driven winds. As a caveat, we mention that at earlier epochs these galaxies likely underwent more active evolutionary phases characterized by gas-rich (minor) mergers, which possibly brought a significant supply of cold gas, thereby triggering starburst driven winds. These starburst driven winds at high redshifts presumably played a role in lifting a certain amount of hot gas to larger radii, as well as enriching the hot gas originating from infall.  

From the derived upper limits on the gas mass and the lower limits on the cooling time, we obtain upper limits on the cooling rates as $\dot M=M_{\rm{gas}} / t_{\rm{cool}}$. Using this estimate we probe if the cooling of the hot corona can balance the ongoing star-formation in the disk. We find that the combined upper limits on the cooling rate in the $(0.05-0.30)r_{200}$ exceed or are comparable to the observed star-formation rates for all galaxies in our sample. Additionally, the cooling of the hot gas in the innermost regions ($<0.05r_{200}$) may also add a contribution, since this gas is expected to be denser, and hence has a shorter cooling time. This implies that cooling of the gas in the corona may be able to provide the cold gas required for the ongoing star-formation. We note that this result is in contrast with that observed for NGC~1961 and NGC~6753, which systems host significantly hotter gas with cooling times that exceed the Hubble-time. 

Although the \textit{Chandra} observations of NGC~266 did not reveal a statistically significant detection of a hot gaseous corona, this result is not in conflict with our earlier detection based on \textit{ROSAT} observations. Specifically, within the $(0.05-0.15)r_{\rm 200}$ region, the observed \textit{ROSAT} count rate was  $\sim3.4\times10^{-3} \ \rm{s^{-1}}$ \citep{bogdan13b}. As opposed to this, the count rate detected by \textit{Chandra} is $\sim1.1\times10^{-3} \ \rm{s^{-1}}$. Taking into account the effective area of the \textit{ROSAT} PSPC and the \textit{Chandra} ACIS-I detectors, we conclude that the factor of $\sim3$ difference in the count rates may be due to the low temperature of the hot gas. Specifically, our observational results are consistent with a picture, in which the hot gas corona surrounding NGC~266 has a temperature of $kT\sim0.15$ keV. Thus, the non-detection of the hot corona around NGC~266 in the deep \textit{Chandra} observations is likely due to the combined effects of the low ACIS-I effective area below $0.5$ keV energy and the low gas temperature.

Since in this work -- in the absence of statistically significant detections -- we assumed a spectral model to derive upper/lower limits on various parameters of the hot gas, we briefly overview the uncertainties that may be associated with the chosen spectral model.  A notable source of uncertainty arises from the unknown gas temperatures. For example, if the gas temperature is factor of two lower/higher than the assumed $0.2$ keV, that would imply a factor of about two higher/lower counts-to-flux conversion for the ACIS-S detector and factor of about three difference for galaxies observed with the ACIS-I detector. Additionally, the unknown metal abundances may also significantly influence the conversion. For example, if the metal abundance are $5$ times higher/lower than the assumed $0.1$ Solar abundance, that would yield a factor of $\sim2$ increase/decrease in the gas densities and gas masses. 

Since in the present study we could not identify a so-far unexplored hot corona, the number of galaxies with detected coronae remains four: NGC~1961, UGC12591, NGC~6753, and NGC~266 \citep{anderson11,dai12,bogdan13a,bogdan13b}. The difficulty in identifying hot coronae around spiral galaxies with present-day telescopes has two major sources. First, the detectable X-ray surface brightness of the coronae is low due to their relatively faint but extended nature. As a result, the observed signal-to-noise ratios, particularly for lower mass galaxies, do not allow to unambiguously detect these coronae. Second, as pointed out by Illustris, the gas temperatures rapidly drops for galaxies with  $M_{\rm \star}\lesssim3\times10^{11} \ \rm{M_{\odot}}$.  However, \textit{Chandra} and \textit{XMM-Newton} have relatively low effective area at energies below $\sim0.5$ keV, and hence are not sufficiently sensitive to thermal emission with $\sim0.1-0.2$ keV. Additionally, for such low temperatures the hot gaseous emission is prone to absorption, which in turn can drastically reduce the observed X-ray flux. The combination of these effects makes it difficult to detect hot coronae around lower mass spiral galaxies, particularly at radii exceeding $\sim0.15r_{\rm 200}$.

\section{Conclusions}
In this paper, we studied the hot gaseous coronae around 8 normal spiral galaxies,  whose stellar mass is in the range of $(0.7-2.0)\times10^{11} \ \rm{M_{\odot}}$, using \textit{Chandra} X-ray observations. To confront the observations with galaxy formation models, we utilized the hydrodynamical cosmological simulation, Illustris. Our results are summarized below:

\begin{enumerate}
\item We did not detect a statistically significant hot corona around any of our sample galaxies in the  $(0.05-0.15)r_{\rm 200}$ or $(0.15-0.30)r_{\rm 200}$ radial ranges. However, for NGC~266 we identified the presence of a moderately significant ($\sim2\sigma$) extended X-ray emitting component in the $(0.05-0.15)r_{\rm 200}$ region, which emission may originate from hot ionized gas. This results is in agreement with our earlier detection based on \textit{ROSAT} data \citep{bogdan13b}. 

\item In the absence of statistically significant detections, we derived $3\sigma$ limits on the X-ray luminosity, the confined gas mass, the  electron densities, and the cooling times. To derive these limits, we assumed that the hot gas is spherically symmetric, has a temperature of $kT=0.2$ keV and metal abundances of $0.1$ Solar. 

\item The derived upper/lower limits, complemented with the gas parameters measured for NGC~1961 and NGC~6753,  were confronted with results of the Illustris Simulation. We concluded that the observed $3\sigma$ upper limits on the X-ray luminosity and hot gas mass exceed or are at the upper end of the model predictions. For NGC~1961 and NGC~6753, we found that the gas temperatures, metal abundances, and electron density profiles broadly agree with the model predictions. These results imply that the observed properties of X-ray coronae around spirals are broadly consistent with those predicted by Illustris. However, for galaxies with massive black holes (mostly giant ellipticals) Illustris predicts overly powerful radio-mode feedback, which in turn results in under luminous X-ray coronae in these systems. 
\end{enumerate}

\smallskip

\begin{small}
\noindent
\textit{Acknowledgements.}
We thank the anonymous referee for his/her constructive comments. We thank Volker Springel, Debora Sijacki, and Dylan Nelson for helpful discussions. This research has made use of software provided by the Chandra X-Ray Center in the application packages CIAO. This work has made use of the NASA/IPAC Extragalactic Database, which is operated by the Jet Propulsion Laboratory, California Institute of Technology, under contract with NASA. We acknowledge the usage of the HyperLeda database (http://leda.univ-lyon1.fr).  \'AB acknowledges support provided by NASA through Einstein Postdoctoral Fellowship awarded by the CXC, which is operated by the SAO for NASA under contract NAS8-03060. WF and CJ acknowledge support from the Smithsonian Institution. LH acknowledges support from NASA grant NNX12AC67G and NSF grant AST-1312095.
\end{small}

\end{document}